\documentclass[twocolumn,final,prb,showpacs,superscriptaddress,amsmath,amssymb,amsfonts,floatfix,epsfig]{revtex4-1}
\usepackage{graphicx}
\usepackage{amsfonts}
\usepackage{amsmath}
\usepackage{amssymb}
\usepackage{epsfig}
\usepackage{multirow}
\usepackage[colorlinks,breaklinks=true,bookmarks=false,citecolor=blue,linkcolor=red,urlcolor=blue]{hyperref}
\newcommand{\nasb}{Na$_3$Cu$_2$SbO$_6$}
\newcommand{\nate}{Na$_2$Cu$_2$TeO$_6$}

\begin{document}

\title{Microscopic magnetic modeling for the $S$\,=\,$\frac12$ alternating chain compounds Na$_3$Cu$_2$SbO$_6$ and Na$_2$Cu$_2$TeO$_6$}

\author{M. Schmitt}
\affiliation{Max-Planck-Institut f\"ur Chemische Physik fester Stoffe, N\"othnitzer Str.\ 40, 01187 Dresden, Germany}
\affiliation{Fachbereich Physik, Technische Universit\"at Kaiserslautern, 67663 Kaiserslautern, Germany}
\author{O. Janson}
\affiliation{Max-Planck-Institut f\"ur Chemische Physik fester Stoffe, N\"othnitzer Str.\ 40, 01187 Dresden, Germany}
\affiliation{National Institute of Chemical Physics and Biophysics, 12618 Tallinn, Estonia}
\author{S. Golbs}
\affiliation{Max-Planck-Institut f\"ur Chemische Physik fester Stoffe, N\"othnitzer Str.\ 40, 01187 Dresden, Germany}
\author{M. Schmidt}
\affiliation{Max-Planck-Institut f\"ur Chemische Physik fester Stoffe, N\"othnitzer Str.\ 40, 01187 Dresden, Germany}
\author{W. Schnelle}
\affiliation{Max-Planck-Institut f\"ur Chemische Physik fester Stoffe, N\"othnitzer Str.\ 40, 01187 Dresden, Germany}
\author{J. Richter}
\affiliation{Institut f\"{u}r Theoretische Physik, Universit\"{a}t Magdeburg, 39016 Magdeburg, Germany}
\author{H. Rosner}
\email{rosner@cpfs.mpg.de}
\affiliation{Max-Planck-Institut f\"ur Chemische Physik fester Stoffe, N\"othnitzer Str.\ 40, 01187 Dresden, Germany}

\date{\today}
\begin{abstract}
The spin-1/2 alternating Heisenberg chain system Na$_3$Cu$_2$SbO$_6$
features two relevant exchange couplings: $J_{1a}$ within the structural
Cu$_2$O$_6$ dimers and $J_{1b}$ between the dimers.  Motivated by the controversially
discussed nature of $J_{1a}$, we perform extensive density-functional-theory
(DFT) calculations, including DFT+$U$ and hybrid functionals.  Fits to the
experimental magnetic susceptibility using high-temperature series
expansions and quantum Monte Carlo simulations yield the optimal parameters
$J_{1a}\!=\!-217$\,K and $J_{1b}\!=\!174$\,K with the alternation ratio
$\alpha=J_{1a}/J_{1b}\simeq-1.25$.  For the closely related system
Na$_2$Cu$_2$TeO$_6$, DFT yields substantially enhanced $J_{1b}$, but weaker
$J_{1a}$. The comparative analysis renders the buckling of the chains as
the key parameter altering the magnetic coupling regime.  Numerical
simulation of the dispersion relations of the alternating chain model
clarify why both antiferromagnetic and ferrromagnetic $J_{1a}$ can
reproduce the experimental magnetic susceptibility data.
\end{abstract}

\pacs{}

\maketitle

\section{Introduction} 
The vibrant research on magnetic insulators keeps on delivering new examples
of exotic magnetic behaviors and unusual magnetic ground states
(GSs).\cite{lee08, balents10} Two prominent examples are the spin-liquid
system herbertsmithite Cu$_3$Zn(OH)$_6$Cl$_2$, featuring a kagome lattice of
$S\!=\!1/2$ spins,\cite{herb} or the recently discovered
Ba$_3$CuSb$_2$O$_9$, where the magnetism is likely entangled with the
dynamical Jahn-Teller distortion.\cite{na3cusb2o9}

Cuprates are a particularly promising playground to study low-dimensional
magnetism, since they often combine the quantum spin $S\!=\!1/2$  ensured by
the Cu $3d^9$ electron configuration and the low-dimensionality of the
underlying magnetic model.  The latter is ensued by the unique variety of
lattice topologies realized in cuprates, which includes geometrically
frustrated lattices, where quantum fluctuations are additionally enhanced by
the competing magnetic interactions.

The simplest example of a quantum GS that lacks a classical analog is the
quantum-mechanical singlet.  Such a GS is found experimentally, e.g., in
CsV$_2$O$_5$ (Ref.~\onlinecite{csv2o5_isobe96, *csv2o5_valenti02,
*csv2o5_saul11}), CuTe$_2$O$_5$ (Ref.~\onlinecite{cuteO_deisenhofer06,
*cuteo_das08, *cuteo_ushakov09}), CaCuGe$_2$O$_6$
(Ref.~\onlinecite{cacuge2O6_sasago95, *cacuge2o6_valenti02}), and
Cu$_2$(PO$_3$)$_2$CH$_2$ (Ref.~\onlinecite{cu2po32ch2}).  All these compounds
feature pairs of strongly coupled spins (magnetic dimers).  An isolated dimer
is an archetypical two-level quantum system, which can be solved analytically.

Compounds with sizable couplings between the dimers can exhibit diverse
behaviors.  For instance, the non-frustrated\cite{mazurenko2013} spin lattice of the Han purple
BaCuSi$_2$O$_6$ is favorable for propagation of triplet excitations, promoting
a Bose-Einstein condensation of magnons, experimentally observed in the
magnetic field range between 23.5 and 49\,T.\cite{bacusi2O6_jaime06,
*bacusi2o6_sebastian06, *bacusi2o6_kraemer07} In contrast, SrCu$_2$(BO$_3$)$_2$
features strongly frustrated interdimer couplings that give rise to a
fascinating variety of magnetization plateaus.\cite{ *[{ }] [{, and references
therein}] SCBO_NMR_review} The remarkable difference between the behavior of
BaCuSi$_2$O$_6$ and SrCu$_2$(BO$_3$)$_2$ is governed by the difference in the
magnetic couplings that constitute the respective spin model.  Thus, the
precise information on the underlying spin model is crucial for
understanding the magnetic properties.  

An evaluation of the microscopic magnetic model can be performed in different
ways.  The basic features of the spin lattice can be often conceived by
applying empirical rules, such as the Goodenough--Kanamori rules.\cite{gka_1,
*gka_2} Then, the resulting qualitative model is parameterized by fitting
its respective free parameters to the experiment. The main challenge is the limited amount of the
available experimental data that may not suffice for a unique and justified fitting of the
model-specific free parameters. Thus, such a phenomenological approach is generally
insecure against ambiguous solutions.

Microscopic modeling based on density-functional theory (DFT) calculations
is an alternative solution. Such calculations require no experimental
information beyond the crystal structure, and in contrast to the
phenomenological method, provide a microscopic insight. A straightforward
application of the DFT is impeded by the fact that cuprates are strongly
correlated materials. Hence the effective one-electron approach of DFT generally fails
to reproduce their insulating electronic GS.\cite{HTSC_Pickett} This
shortcoming can be mended in alternative calculational schemes, such as
DFT+$U$ or hybrid functionals, yet these methods are not parameter-free.
Often, these parameters sensitively depend on the fine structural details of
the system under investigation.

The low-dimensional $S\!=\!1/2$ Heisenberg compound \nasb\ is an
instructive example that demonstrates the performance and the
limitations of the phenomenological as well as the microscopic approach.
This compound was initially described as a distorted honeycomb lattice,
owing to the hexagonal arrangement of the Cu atoms in the crystal
structure.\cite{miura06} However, this purely geometrical analysis
neglects the key ingredients of the magnetic superexchange, such as the
orientation and the spatial extent of the magnetically active orbitals.
Indeed, as pointed out by the authors of Ref.~\onlinecite{miura06}, the
orientation of the Cu $3d_{x^2-y^2}$ orbitals readily accentuates the
chains formed by structural dimers and hints at two relevant magnetic
couplings: $J_{\text{1a}}$ within the structural dimers and
$J_{\text{1b}}$ between the dimers (Fig.~\ref{F-str}), leading to the
quasi-1D Heisenberg chain model with alternating nearest-neighbor
couplings.

Thermodynamical measurements confirmed the quasi-1D character of
the spin model,\cite{miura06, derakhshan07} yet no agreement was found for the
sign of the intradimer coupling $J_{\text{1a}}$:
Refs.~\onlinecite{miura06} and \onlinecite{derakhshan07} vouch for a
ferromagnetic (FM) and antiferromagnetic (AFM) exchange, respectively.
The sign of $J_{\text{1a}}$ basically governs the magnetic GS: the
AFM-AFM solution is a disordered dimer state, while the GS of an FM-AFM
chain is adiabatically connected to the Haldane phase with nontrivial
topology and sizable string order parameter.\cite{AHC_phase_diagram}
Therefore, for the magnetic GS, the sign of $J_{\text{1a}}$ is of
crucial importance.

Notably, even DFT studies do not concur with each other:
Ref.~\onlinecite{derakhshan07} reports AFM $J_{\text{1a}}$, while an
alternative DFT-based method in Ref.~\onlinecite{koo08} yields FM
coupling.  To resolve the controversy on the sign of $J_{\text{1a}}$,
the authors of Ref.~\onlinecite{miura08} performed inelastic neutron
scattering (INS) experiments on single crystals of \nasb.  The resulting
values for the exchange couplings ($J_{\text{1a}}$\,=\,$-145$\,K and
$J_{\text{1b}}$\,=\,161\,K) clearly indicate the FM-AFM chain
scenario.  Still, the origin of ambiguous solutions in earlier
experimental as well as in DFT studies has not been sufficiently
clarified.

In our combined experimental and theoretical study, we evaluate the
magnetic model for Na$_3$Cu$_2$SbO$_6$ and its Te sibling
Na$_2$Cu$_2$TeO$_6$ (Ref.~\onlinecite{xu05}) using extensive DFT
calculations and investigate how the magnetic GS is affected by the
structural distortion within the chains.  By comparing our DFT results
to the earlier studies, we explain the origin of ambiguous
parameterizations of DFT-based spin models in both compounds.
Simulations of the momentum-resolved spectrum for our microscopic model
reveal excellent agreement with the INS experiments
(Ref.~\onlinecite{miura08}) and enlighten the ambiguity of AFM--AFM and
FM-AFM solutions inferred from the thermodynamical measurements.

This paper is organized as follows. The used experimental as well as
computational methods are described in Sec.~\ref{S-methods}. The details
of the crystal structures of \nasb\ and \nate\ are discussed in
Sec.~\ref{S-str}.  In Sec.~\ref{S-res}, we present our magnetic
susceptibility measurements and extensive DFT calculations.
Peculiarities of the excitation spectrum of the Heisenberg chain model
is discussed Sec.~\ref{S-spectrum}.  Finally, a summary and a short
outlook are given in Sec.~\ref{S-sum}.

\section{\label{S-methods}Methods}

\paragraph*{Synthesis and sample characterization} Polycrystalline
samples of Na$_3$Cu$_2$SbO$_6$ were prepared by solid state reaction. A
stoichiometric amount of Na$_2$CO$_3$ (Chempur, 99.9+\%), Sb$_2$O$_5$
(99.999\%, Alfa Aesar) and CuCO$_3$$\cdot$Cu(OH)$_2$ (Chempur) was
thoroughly mixed. The homogeneous powder was pressed into a platinum
crucible and annealed at 1273\,K for two weeks in air. Finally the
crucible was taken out of the furnace at 1273\,K and cooled down to room
temperature in air. 

For magnetic measurements, the powder sample was pressed into a pellet
and heated again at 973\,K in a platinum boat for several days.  The
green powder was identified and characterized by powder x-ray
diffraction using a high-resolution Guinier camera with Cu K$_{\alpha}$
radiation. The determined lattice parameters $a\!=\!5.676$\,\AA,
$b\!=\!8.860$\,\AA, $c\!=\!5.833$\,\AA\ and $\beta\!=\!113.33^{\circ}$
are in good agreement with Ref.~\onlinecite{smirnova05}.

To control the oxygen content in the sample at different stages of the
thermal treatment, we performed coulometric titration of the samples
using a commercial OXYLYT device.  We found that the maximal oxygen
content (close to the stoichiometric \nasb) is attained right after the
thermal treatment at 973\,K (Fig.~S3 in~Ref.~\onlinecite{suppl}).
However, a subsequent storage at room temperature and in air leads to a
reduction of the oxygen content.  This effect can be seen in the
magnetic susceptibility by the increased amount of Curie impurity
(Fig.~S4 in~Ref.~\onlinecite{suppl}).  Therefore, for thermodynamic
measurements, we use ``fresh'' samples (i.e., we performed measurement
right after the thermal treatment) that feature smallest impurity
contribution.  Magnetic susceptibility $\chi(T)$ of \nasb\ was measured
using a SQUID magnetometer (MPMS, Quantum Design) in a magnetic field of
0.04\,T.

\paragraph*{DFT calculations} For the electronic structure calculations,
the full-potential local-orbital code FPLO (version
\textsc{fplo8.50-32}) within the local (spin) density approximation
(L(S)DA) was used.\cite{Koep1} In the scalar relativistic calculations
the exchange and correlation potential of Perdew and Wang has been
applied.\cite{PerdW} The accuracy with respect to the $k$-mesh has been
carefully checked.

The LDA band structure has been mapped onto an effective one-orbital
tight binding (TB) model based on Cu-site centered Wannier functions
(WF).  The strong Coulomb repulsion of the Cu $3d$ orbitals was
considered by mapping the TB model onto a Hubbard model. In the strongly
correlated limit and at half-filling, the lowest lying (magnetic)
excitations can be described by  a Heisenberg model with
$J^{\text{AFM}}_{ij}\!=\!4t_{ij}^2/U_{\text{eff}}$ for the
antiferromagnetic part of the exchange.  Spin-polarized LSDA+$U$
supercell calculations were performed  using two limiting cases for the
double counting correction (DCC): the around-mean-field (AMF) and the
atomic limit (AL, also called the fully localized limit).  We varied the
on-site Coulomb repulsion $U_{3d}$ in the physically relevant range
(4--8\,eV in AMF and 5--9\,eV in AL), keeping the on-site exchange
$J_{3d}\!=\!1$\,eV.

The partial Na$_{3-x}$ occupancy and the Sb$_x$/Te$_{1-x}$ substitution were
modeled using the virtual crystal approximation (VCA).\cite{kasinathan2012}

HSE06 (Ref.~\onlinecite{HSE03, *HSE06}) hybrid functional calculations were performed using the
pseudopotential code \textsc{vasp-5.2},\cite{vasp1, *vasp2} employing the
basis set of projector-augmented waves.  The default admixture of the Fock
exchange (25\%) was adopted. We used the primitive unit cell with 2 Cu atoms
and a 6$\times$6$\times$6 $k$-mesh with the NKRED=3 flag.

\paragraph*{Simulations and fits to the experiment} We used
the high-temperature series expansion (HTSE) to a Heisenberg chain with alternating
nearest-neighbor couplings $J_{\text{1a}}$ and $J_{\text{1b}}$. For the case
of AFM couplings, the parameterization is given in
$\alpha\equiv|J_{\text{1a}}|/J_{\text{1b}}$ in Table II of
Ref.~\onlinecite{HC_AHC_Johnston}; the parameters for the case of FM
$J_{\text{1a}}$ are provided in Ref.~\onlinecite{AHC_HTSE}.  Quantum Monte Carlo
simulations were performed using the \textsc{loop} algorithm\cite{looper} from
the \texttt{ALPS} package.\cite{ALPS} To evaluate the reduced magnetic
susceptibility, we used 50\,000 loops for thermalization and 500\,000 loops
after thermalization for chains of $N\!=\!120$ spins $S\!=\!1/2$ using
periodic boundary conditions.  Exact (Lanczos) diagonalization of the Heisenberg
Hamiltonians was performed using \textsc{spinpack}.\cite{spinpack} The
lowest-lying $S^z$\,=\,0, $S^z$\,=\,1 and $S^z$\,=\,2 excitations were computed
for $N$\,=\,32 sites chains of $S$\,=\,1/2 using periodic boundary conditions.

\section{\label{S-str}Crystal structure}

\begin{figure}
\includegraphics[width=8.7cm]{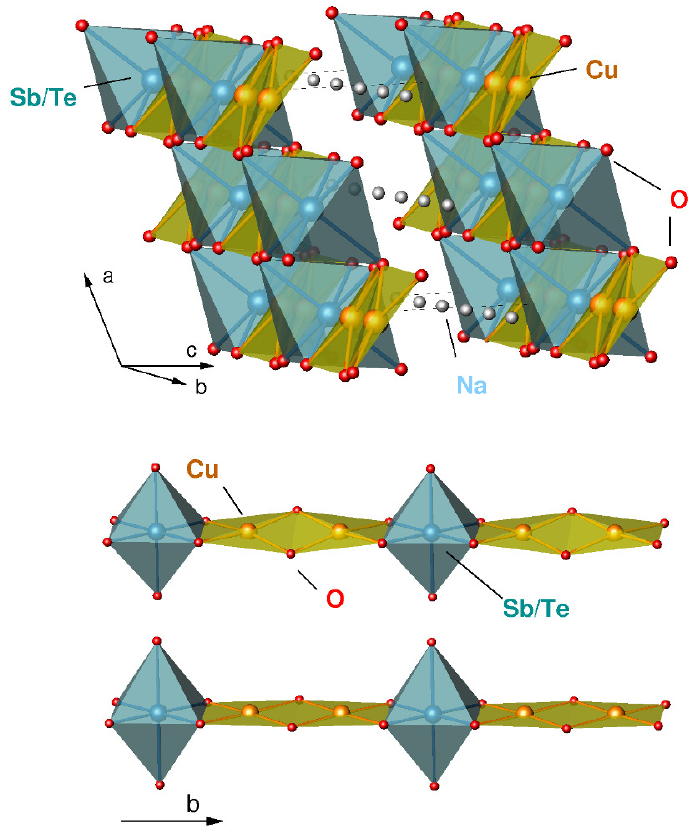}
\caption{\label{F-str} (Color online) Top: crystal structure of
Na$_3$Cu$_2$SbO$_6$. The basic elements are CuO$_4$ plaquettes
and SbO$_6$ octahedra. Bottom: segments of the structural chains of
Cu$_2$O$_6$ dimers for the experimental distorted plaquette geometry
(upper panel) and an ideal planar arrangement of the Cu$_2$O$_6$ units
in the fictitious structures (lower panel).} \end{figure} 

The monoclinic (space group $C2/c$) crystal structure of Na$_3$Cu$_2$SbO$_6$
(Ref.~\onlinecite{smirnova05}) features pairs of slightly distorted,
edge-shared CuO$_4$ plaquettes forming structural dimers with the Cu--O--Cu
bonding angle of 95\,$^\circ$. The dimers are connected by the equatorial plane
of SbO$_6$ octahedra and form chains running along the $b$ axis
(Fig.~\ref{F-str}, bottom). The apical O atoms of the SbO$_6$ octahedra mediate
connections to the next Cu$_2$O$_6$ dimer chain.  In this way, the magnetic
layers, separated by Na atoms, are formed (Fig.~\ref{F-str}, top).

The crystal structure of Na$_2$Cu$_2$TeO$_6$ (Ref.~\onlinecite{xu05}) features
a similar motif, with the reduced number of Na atoms between the layers, to
keep the charge balance.  In addition, the smaller size of Te$^{6+}$ compared
to Sb$^{5+}$ gives rise to a stronger distortion of the Cu$_2$O$_6$ dimer
chains in Na$_2$Cu$_2$TeO$_6$.  To investigate the influence of this
distortion, we also computed fictitious structures with idealized planar arrangements of
the Cu$_2$O$_6$ units (Fig.~\ref{F-str} bottom, lower panel).

\section{\label{S-res}Results}

\subsection{Magnetic susceptibility}
Above 200\,K, the magnetic susceptibility of \nasb\ fits reasonably to the
Curie-Weiss law with $C$\,=\,0.442\,emu\,K(mol\,Cu)$^{-1}$ and the
antiferromagnetic Weiss temperature $\theta_{\text{CW}}$\,=\,60$\pm$10\,K. The
effective magnetic moment amounts to
$\mu_{\text{eff}}\!\simeq\!1.88$\,$\mu_{\text{B}}$, slightly exceeding the
spin-only value for $S$\,=\,1/2 (1.73\,$\mu_{\text{B}}$). The resulting value
of the Lande factor $g$\,=\,2.17 is typical for Cu$^{2+}$ compounds.  At
lower temperatures, antiferromagnetic correlations give rise to a broad maximum
in the magnetic susceptibility around $T_{\text{max}}$\,=\,96\,K. The
low-temperature upturn below 17\,K is likely caused by defects, typical for
powder samples of quasi-1D magnets $[$e.g., Sr$_2$Cu(PO$_4$)$_2$ from
Ref.~\onlinecite{HC_Sr2CuPO42_Ba2CuPO42_str_chiT_DTA} or (NO)Cu(NO$_3$)$_3$
from Ref.~\onlinecite{HC_NOCuNO33_chiT_CpT_ESR_simul}$]$, since already a single
defect terminates the spin chain.

\begin{figure}[tb]
\includegraphics[width=8.6cm]{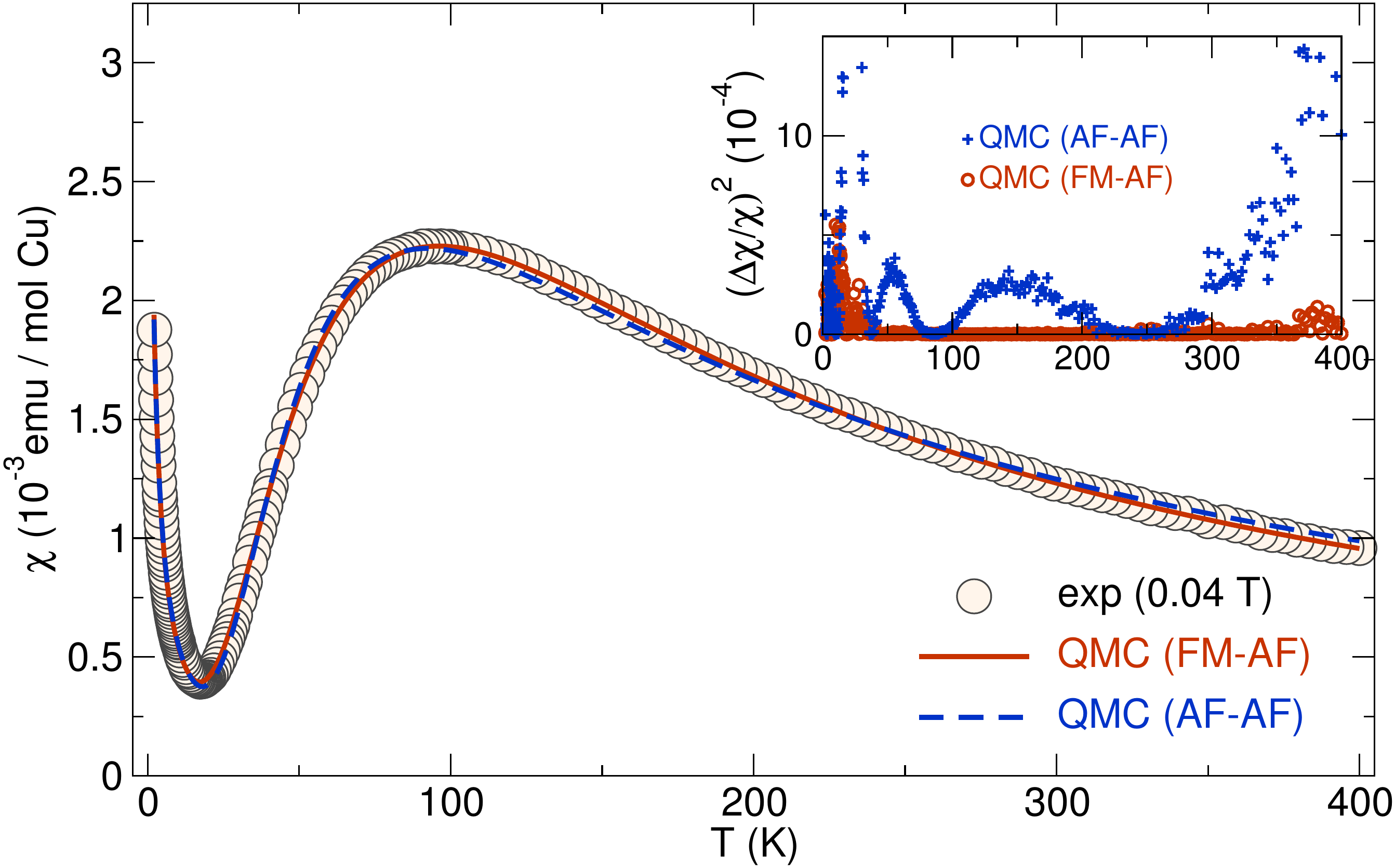}
\caption{\label{F-chiT}(Color online) Experimental (exp) magnetic
susceptibility of \nasb\ (circles) and the quantum Monte Carlo (QMC) fits for
the FM-AFM and the AFM-AFM solution of the alternating Heisenberg chain (AHC)
model. Inset: difference curves emphasize the excellence of the FM-AFM
solution.} \end{figure}

We briefly compare our susceptibility measurements with the published data.
The Curie-Weiss fit from Ref.~\onlinecite{derakhshan07} yields a similar
$\theta_{\text{CW}}$\,=\,55\,K, but their $g$\,=\,2.33 exceeds our
estimate. This discrepancy likely originates from the difference in the
magnetic field (0.1\,T versus 0.04\,T in our work) as well as different
temperature ranges used for the fitting.  Unfortunately, the authors of
Ref.~\onlinecite{miura06} do not provide the values of $\theta_{\text{CW}}$ and
$g$, but a Curie-Weiss fit to their data yields
$\theta_{\text{CW}}$\,$\simeq$\,49\,K and $g$\,$\simeq$\,2.10, in good
agreement with our findings. A bare comparison of the absolute values of
$\chi(T_{\text{max}})$ (Table~\ref{T-cw}) reveals sizable deviations of
the $\chi(T)$ data from Ref.~\onlinecite{derakhshan07} compared to the other
two data sets.

\begin{table}[tb]
\caption{\label{T-cw} \nasb: the Curie-Weiss temperature $\theta_{\text{CW}}$ (in K) and
the $g$-factor evaluated using the Curie-Weiss fit for $T\!\geq\!200$\,K, as well as
the experimental position $T_{\text{max}}$ (in K) of the susceptibility maximum
and its abosolute value $\chi(T_{\text{max}})$ $[$in emu\,(mol\,Cu)$^{-1}]$.}
\begin{ruledtabular}
\begin{tabular}{r c c c c} 
data source & $\theta_{\text{CW}}$ & $g$ & $T_{\text{max}}$ & $\chi(T_{\text{max}})$ \\ \hline
this study & 60 & 2.17 & 96 & 2.2$\cdot10^{-3}$\\
data from {Ref.~\onlinecite{miura06}} & 49 & 2.10 & 95 & 2.3$\cdot10^{-3}$ \\
{Ref.~\onlinecite{derakhshan07}} & 55 & 2.33 & 90 & 1.7$\cdot10^{-3}$\\
\end{tabular}
\end{ruledtabular}
\end{table}

For a more elaborate analysis, we adopt the AHC model and search for
solutions that agree with the experimental $\chi(T)$ curve. To this end, we
perform HTSE considering the physically different scenarios: both
$J_{\text{1a}}$ and $J_{\text{1b}}$ couplings are AFM (``AFM--AFM'') and
$J_{\text{1a}}$ is FM (``FM--AFM''). The corresponding HTSE coefficients for
the two cases can be found in Refs.~\onlinecite{HC_AHC_Johnston} and
\onlinecite{AHC_HTSE}, respectively. In both cases, we obtain a solution
(first row of Table~\ref{T-chiT}) which conforms to the experimental data.

\begin{table}[tb]
\caption{\label{T-chiT} High-temperature series expansion (HTSE) and quantum
Monte Carlo (QMC) fits to the experimental $\chi(T)$ data for \nasb.
Results of different studies implying a ferromagnetic and an antiferromagnetic
coupling (FM-AFM, upper lines) or two inequivalent antiferromagnetic couplings
(AFM-AFM, lower lines) are shown: exchange couplings $J_{\text{1a}}$ and
$J_{\text{1b}}$ (in K), $g$-factors, temperature-independent terms $\chi_0$
$[$in emu\,/\,(mol\,Cu)$^{-1}]$, and Curie-Weiss impurity contributions
$C^{\text{imp}}$ $[$in K\,emu\,(mol\,Cu)$^{-1}]$ and
${\theta}^{\text{imp}}_{\text{CW}}$ (in K).  ``$-$'' stands for a fitted
quantity, which numerical value is not provided in the respective reference.}
\begin{ruledtabular}
\begin{tabular}{l r r r r r r} 
& $J_{\text{1a}}$ & $J_{\text{1b}}$ & $g$ & $\chi_0$ & $C^{\text{imp}}$  & ${\theta}^{\text{imp}}_{\text{CW}}$ \\ \hline
\multicolumn{7}{c}{HTSE}\\ \hline 
\multirow{2}{*}{this study} & $-207$ & 171 & 2.01 & 3$\times10^{-5}$ & 4.7$\times10^{-3}$ & \\
& 155 & 66 & 2.20 & $3\times10^{-6}$ & 6.1$\times10^{-3}$ & 1.1 \\ \\
\multirow{2}{*}{Ref.~\onlinecite{miura06}} & $-209$ & 165 & 2.01 & $-$ & $-$ & $-$ \\
& 143 & 39 & 2.13 & $-$ & $-$ & $-$ \\ \\
{Ref.~\onlinecite{derakhshan07}} & 160 & 62 & 1.97 & 2.2$\times10^{-4}$ & & $2.3$ \\ \hline
\multicolumn{7}{c}{QMC}\\ \hline
\multirow{2}{*}{this study} & $-$217 & 174 & 2.02 &  9$\times10^{-6}$ & 6$\times10^{-3}$ &
1 \\
& 153 & 61 & 2.19 & 3$\times10^{-6}$ & 6$\times10^{-3}$ & 1.2\\
\end{tabular}
\end{ruledtabular}
\end{table}

Our solution for the FM-AFM case (Table~\ref{T-chiT}, first row) nearly
coincides with the corresponding solution from
Ref.~\onlinecite{miura06} (Table~\ref{T-chiT}, second
row), yielding $\alpha\equiv{J_{\text{1a}}/J_{\text{1b}}}\!\simeq\!-1.25$ and
a considerably smaller $g$-factor of about 2 compared to the value from the
Curie-Weiss fits (2.17). For the AFM-AFM case, we obtain $\alpha\!\simeq\!0.4$
which deviates from the result of
Ref.~\onlinecite{miura06}, but closely resembles the
solution from Ref.~\onlinecite{derakhshan07}
(Table~\ref{T-chiT}, third row). The discrepancy can originate from
different parameterizations used for the HTSE fitting. In particular, the AFM-AFM
solutions in Refs.~\onlinecite{miura06} and
~\onlinecite{derakhshan07} are obtained using the
parametrization from Ref.~\onlinecite{AHC_HTSE_AFAF}. In contrast, we adopt the
coefficients from a more recent and extensive study,\cite{HC_AHC_Johnston}
valid in the whole temperature range measured.

To account for the full temperature range measured, we turn to QMC simulations.
Thus, we adopt the ratios $\alpha\!=\!-1.25$ and $\alpha\!=\!0.40$ from our
HTSE fitting, and calculate the reduced magnetic susceptibility
$\chi^{*}(T/k_BJ)$, which can be fitted to the experimental curve using the
expression: 

\begin{equation}
\chi(T)=
\frac{N_Ag^2{\mu}_{B}^2}{k_BJ}\cdot\chi^{*}\biggl(\frac{T}{k_BJ}\biggl)
+ \frac{C^{\text{imp}}}{T+\theta^{\text{imp}}_{\text{CW}}} + \chi_0,
\end{equation}
where $N_A$ and $k_B$ are the Avogadro and Boltzmann constants, respectively,
$\mu_{B}$ the Bohr magneton, $C^{\text{imp}}$ and
$\theta^{\text{imp}}_{\text{CW}}$ account for impurity/defect contributions,
$\chi_0$ is a temperature-independent term, and
$J=\max\{|J_{\text{1a}}|,J{_\text{1b}}\}$.  Using a least-squares fitting, we
obtain the solutions listed in Table~\ref{T-chiT} (last row) and shown in
Fig.~\ref{F-chiT}. 

The AFM--AFM solution shows sizable deviations at high temperatures and in the
vicinity of the low-temperature upturn (Fig.~\ref{F-chiT}, inset), while the
FM-AFM solution yields an excellent fit to the experimental $\chi(T)$ in the
whole temperature range, making the latter solution more favorable. Still, the
choice is impeded by the following issues. First, the AHC model is a minimal
model for \nasb, which completely neglects interchain couplings and
anisotropies. Second, the $g$-factor of the FM-AFM solution deviates
significantly from the estimate based on the Curie-Weiss fit, while its
counterpart from the AFM-AFM solution shows a better agreement with the Curie-Weiss fit.
Finally, the shape of the $\chi(T)$ curve is affected by oxygen deficiency in
the sample,\cite{suppl} which is difficult to control during the synthesis
process. Therefore, the AFM-AFM solution can not be ruled out using the $\chi(T)$
data, only.

\subsection{Electronic structure and magnetic model}
\begin{figure}
\includegraphics[width=8.7cm]{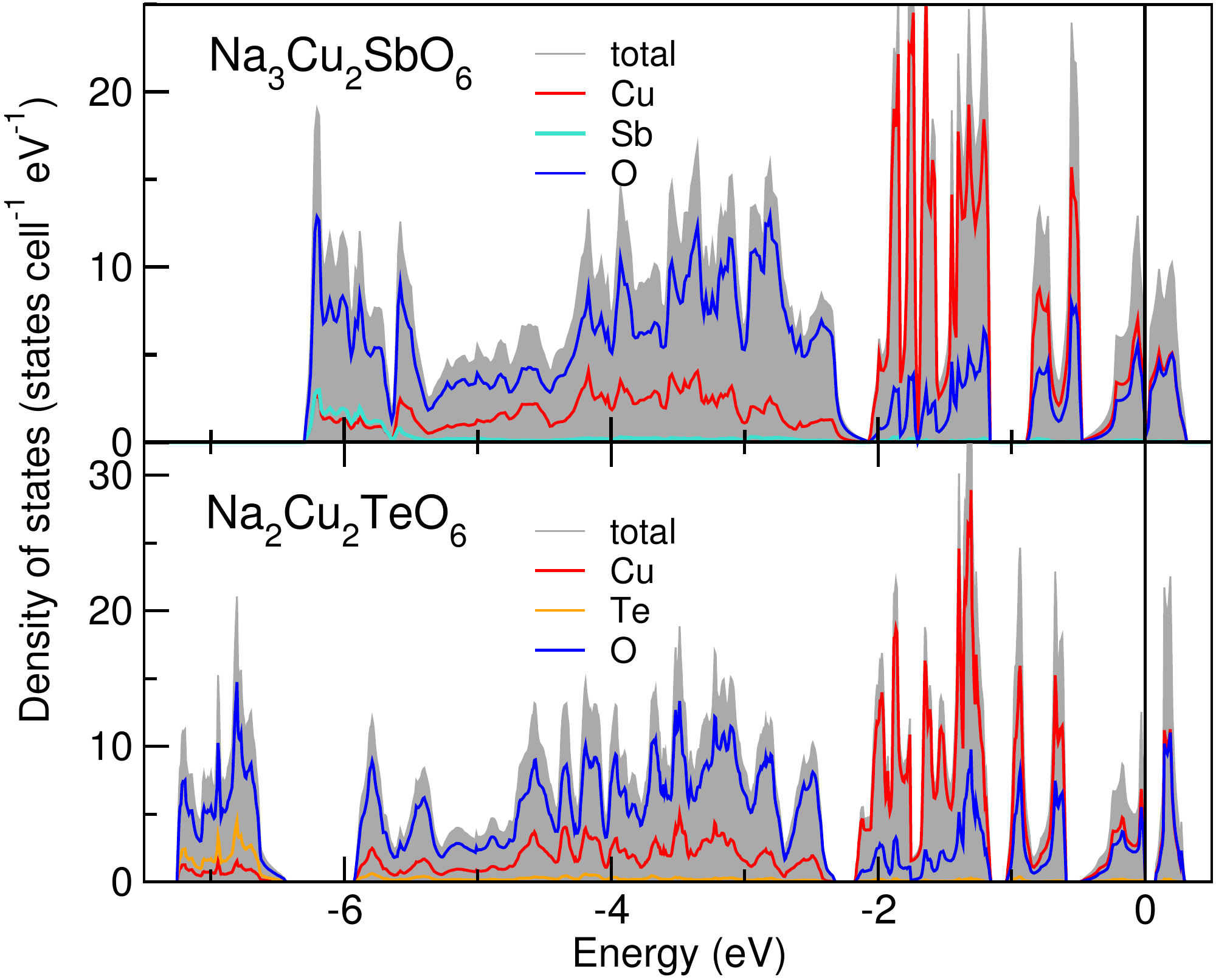}
\caption{\label{dos_atom} 
(Color online) Total and atom-resolved LDA density of states for
Na$_3$Cu$_2$SbO$_6$ (top) and Na$_2$Cu$_2$TeO$_6$ (bottom). 
The contribution of Na states is negligible on this scale (not shown).}
\end{figure}

To resolve the ambiguity between the FM-AFM and AFM-AFM solutions, we perform
microscopic magnetic modeling of Na$_3$Cu$_2$SbO$_6$ and its Te sibling
Na$_2$Cu$_2$TeO$_6$ using DFT calculations.  The valence bands feature similar
band width and are similarly structured in the two compounds, as revealed by
the LDA densities of states (DOS) in Fig.~\ref{dos_atom}.  The DOS is dominated
by Cu and O states down to $-5.5$\,eV and $-6$\,eV for Na$_3$Cu$_2$SbO$_6$ and
Na$_2$Cu$_2$TeO$_6$, respectively. Contributions from Na, Sb and Te are
marginal in this energy range. Only at the lower edge of the valence band, we
find a sizable hybridization of Sb states for Na$_3$Cu$_2$SbO$_6$ centered
around $-6$\,eV. A similar admixture of Te states is observed for
Na$_2$Cu$_2$TeO$_6$, where the additional valence electron of Te compared to Sb
shifts the Cu--O--Te density down by about 1\,eV.
 
\begin{figure}
\includegraphics[width=8.7cm]{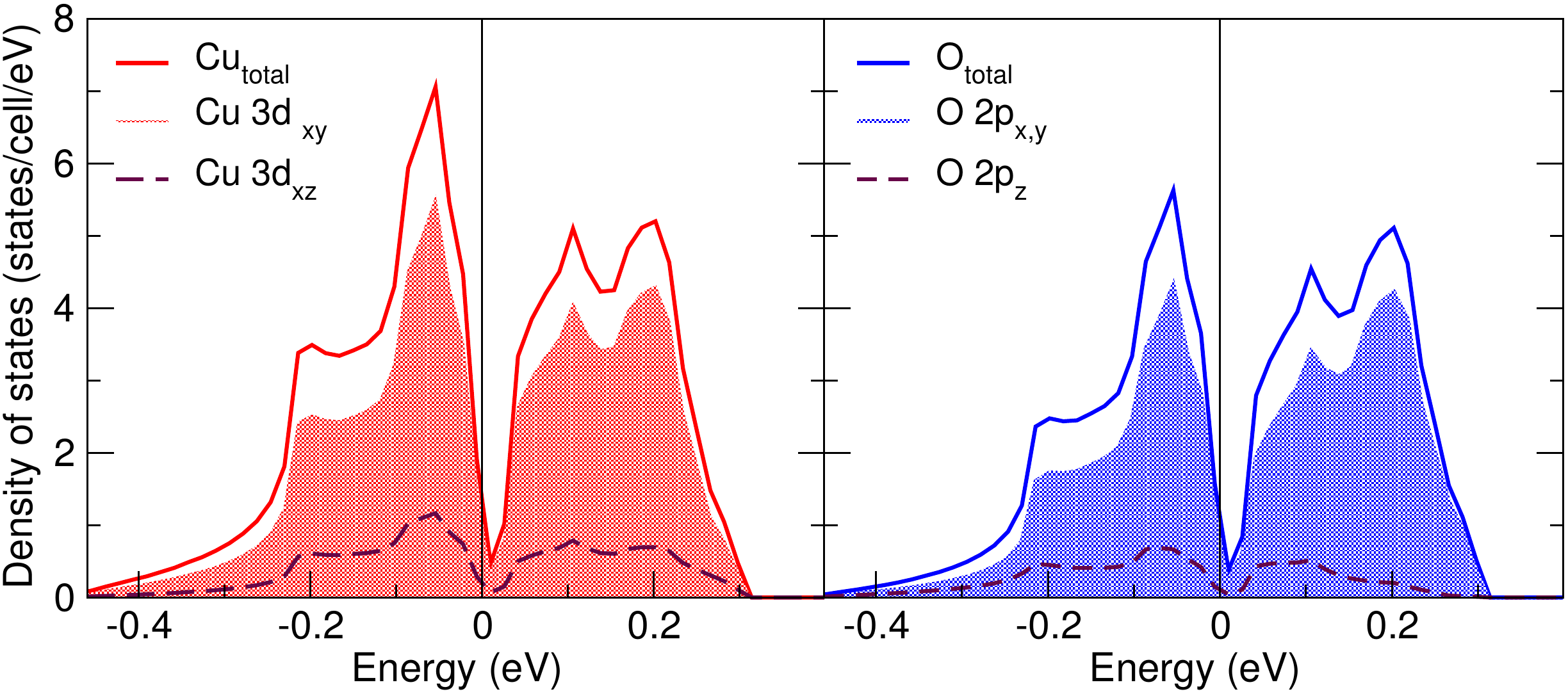}
\caption{\label{dos_orb} 
(Color online) Orbital-resolved LDA density of states for Na$_3$Cu$_2$SbO$_6$. In
the local coordinate system of a CuO$_4$ plaquette, the $x$ axis runs along one of the
Cu--O bonds, while the $z$ axis is perpendicular to the plaquette plane (for
the ideal planar coordination).  The states in the vicinity of the Fermi energy
are dominated by the in-plane Cu $3d$ and O $2p$ states.}
\end{figure}

The LDA band structures for both compounds feature a well-separated density of Cu
and O states centered around the Fermi energy. In the local coordinate system
of a CuO$_4$ plaquette, this density is formed by the anti-bonding
$\sigma$-combination of Cu $3d_{x^2-y^2}$ and O $2p_{\sigma}$ states
($dp\sigma^{*}$ combination).  The orbital-resolved density of states for
Na$_3$Cu$_2$SbO$_6$ is shown in Fig.~\ref{dos_orb}.  Two aspects should be
pointed out.  First, the metallic solution (nonzero DOS at the Fermi energy)
observed in Na$_3$Cu$_2$SbO$_6$, is in contrast with the green color of the
compound, indicative of the insulating behavior.  
Similar, the calculated LDA band gap of 0.06\,eV for Na$_2$Cu$_2$TeO$_6$ (see Fig.~\ref{dos_atom})
is far too small to account for the green color of the powder and originates from dimerization effects.
This drastic
underestimation of the band gap is a well-known shortcoming of the LDA, which
does not account for the strong Coulomb repulsion in the Cu 3$d$ orbitals.  The
missing part of correlation energy will be accounted for by resorting to a
Hubbard model, as well as using DFT+$U$ and hybrid-functional calculations.
Second, the orbital resolved density of states (see
Fig.~\ref{dos_orb}) shows small hybridization with the out-of-plane Cu-O states
due to the distortion of the dimer chains. Since these contributions are small
compared to the pure antibonding $dp\sigma^{*}$ states, the restriction to an
effective TB model is still justified.

To verify the structural input, we relaxed the crystal structures within LDA.
For Na$_3$Cu$_2$SbO$_6$, the relaxation results in a rather small energy gain
of 33\,meV per formula unit (f.\,u.), and the respective changes in the crystal
structure are negligible. In contrast, a relaxation of the atomic coordinates
in Na$_2$Cu$_2$TeO$_6$ lowers the energy by 130\,meV per f.\,u.\ and alters
mainly the chain buckling.  Since the relaxation of Na$_2$Cu$_2$TeO$_6$ affects
the magnetically relevant $dp\sigma^{*}$ states, we evaluated the magnetic
properties for both, the experimental and the relaxed crystal structure.

\begin{figure}
\includegraphics[width=8.7cm]{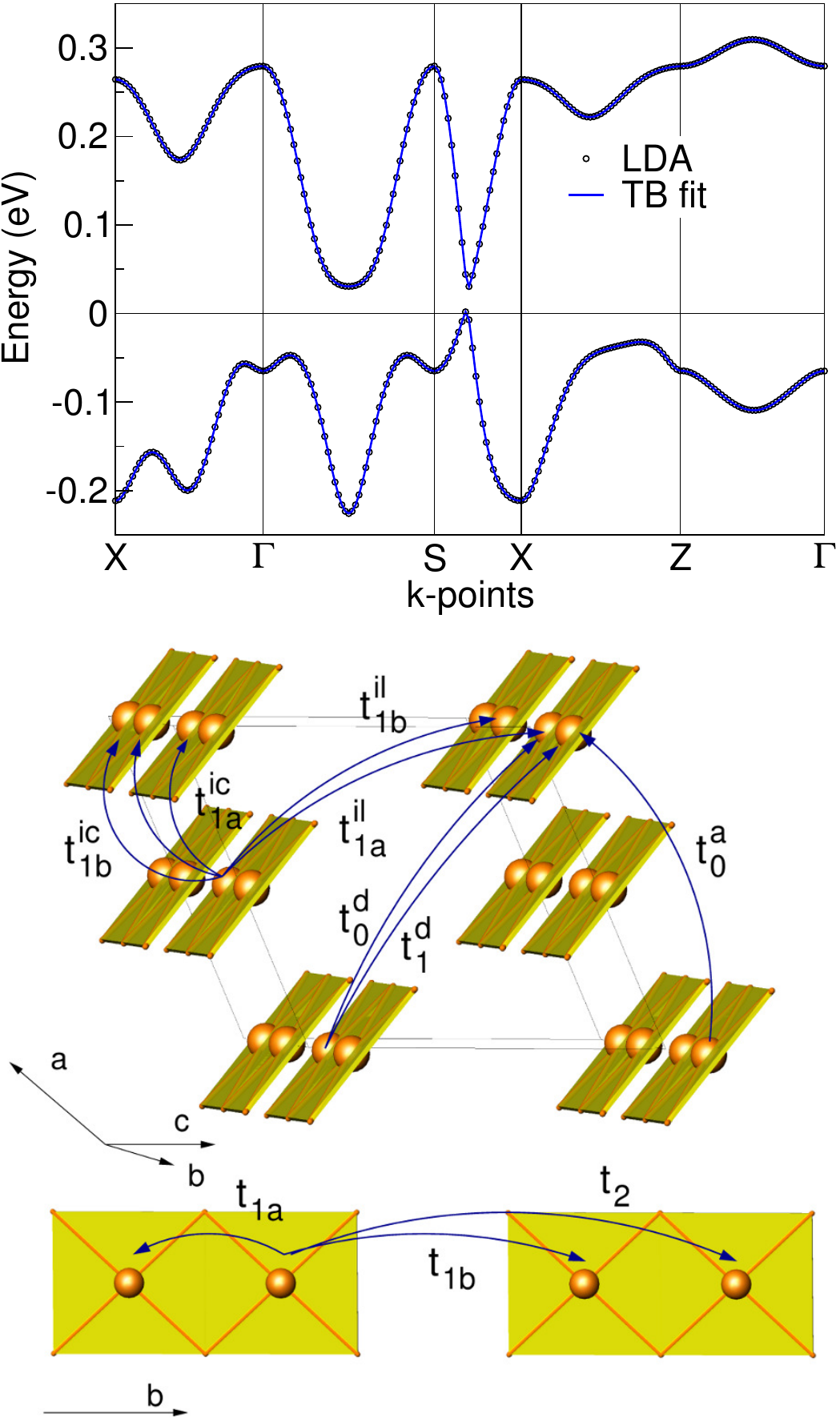}
\caption{\label{tbm} 
(Color online) Top: LDA band structure of Na$_3$Cu$_2$SbO$_6$ and the fit using
the effective one-orbital TB model with ten inequivalent transfer integrals
$t_{ij}$.  Bottom: the superexchange pathways
corresponding to the relevant $t_{ij}$ terms.}
\end{figure}

The transfer integrals $t_{ij}$ (the hopping matrix elements) are evaluated by
a least-squares fit of an effective one-orbital TB model to the two LDA
$dp\sigma^{*}$ bands.  Using 10 inequivalent $t_{ij}$ terms (see the bottom
panel of Fig.~\ref{tbm}, Table~\ref{tab_ts}, and Ref.~\onlinecite{suppl})
we obtain excellent agreement between the TB model and the LDA band
structure. The respective fit for Na$_3$Cu$_2$SbO$_6$ is shown in
Fig.~\ref{tbm} (top). 
 
\begin{table}[htb]
\caption{\label{tab_ts} Relevant ($>$10\,meV) transfer
integrals $t_{ij}$ (in meV) evaluated by fitting the LSDA band structures for
the different structural models: experimental (exp), LSDA-relaxed (relaxed) and
fictitious planar (planar).  For the notation of $t_{ij}$, see Fig.~\ref{tbm} (bottom).}
\begin{ruledtabular}
\begin{tabular}{l ccccccc}
\multicolumn{8}{c}{Na$_3$Cu$_2$SbO$_6$}\\ \hline
$t_i$/meV & $t_{1a}$ & $t_{1b}$ & $t_{2}$ & $t^{ic}_{1a}$& $t^{ic}_{1b}$  &  $t_{0}^{d}$& $t_{0}^{a}$ \\
 exp & {\bf 60.6} & {\bf 127} & 18.2& $-$27.8 & 17.0 & 21.8& 17.4 \\
relaxed & {\bf 68.2} & {\bf 134} & 18.1 & $-$32.3 & 20.6 &20.9 & 19.2  \\
planar exp & {\bf 45.3} & {\bf 119} & 22.4 &  $-$7.8 & 9.4 &  30.1& $-$  \\ 
planar relax.&  {\bf 55.6} & {\bf 125}  &23.8 & $-$9.2& 10.7 & 29.2 & $-$ \\
\hline
\multicolumn{8}{c}{Na$_2$Cu$_2$TeO$_6$}\\ \hline 
$t_i$/meV & $t_{1a}$ & $t_{1b}$ & $t_{2}$ & $t^{ic}_{1a}$& $t^{ic}_{1b}$  &  $t_{0}^{d}$& $t_{0}^{a}$ \\
exp & {\bf 15.6} & {\bf 162} & 16.4& $-$38.5 & 24.7  & 13.7&25.5 \\
relaxed & {\bf 42.5} & {\bf 152} & 17.3 & $-$42.4 & 26.3 &  14.5 &23.1\\
planar exp. &{\bf 27.3} & {\bf 152} & 29.3& $-$12.6 & 12.4 &  25.6& 1.3\\
planar relax. &{\bf 45.2} & {\bf 148} & 30.0& $-$12.8  & 12.7  & 26.0  & $-$\\
\end{tabular}
\end{ruledtabular}
\end{table}

In both systems, the leading coupling is $t_{1b}$, which connects two
neighboring structural dimers: $t_{1b}\!=\!127$\,meV for Na$_3$Cu$_2$SbO$_6$
and $t_{1b}\!=\!162$\,meV for Na$_2$Cu$_2$TeO$_6$,
respectively. The coupling within the structural Cu$_2$O$_6$ dimers
($t_{1a}\!=\!60$\,meV for Na$_3$Cu$_2$SbO$_6$ and $t_{1a}\!=\!16$\,meV for Na$_2$Cu$_2$TeO$_6$) are significantly smaller.  Besides,
several long-range couplings that connect different chains, are comparable to
$t_{1a}$ (Table~\ref{tab_ts} and Ref.~\onlinecite{suppl}).  Subsequent mapping
of the TB model onto a Hubbard model (adopting $U_{\text{eff}}\!=\!4$\,eV) and
a Heisenberg model, yield the following AFM contributions: 
$J^{\text{AFM}}_{1b}\!=\!188$\,K and $J^{\text{AFM}}_{1a}\!=\!43$\,K for
Na$_3$Cu$_2$SbO$_6$ and $J^{\text{AFM}}_{1b}\!=\!305$\,K and
$J^{\text{AFM}}_{1a}\!=\!2$\,K for Na$_2$Cu$_2$TeO$_6$, respectively.

The resulting minimal model is incomplete, since it disregards the FM
contribution to the exchange integrals, which are expected to be especially
large for the $J_{1a}$ coupling within the structural dimers.  To estimate the
total exchange integrals, comprising AFM and FM contributions, we performed
LSDA+$U$ calculations of magnetic supercells.  Mapping the total energies of
different collinear spin arrangements onto a classical Heisenberg model yields
$J_{1a}\!=\!-135\pm20$\,K for Na$_3$Cu$_2$SbO$_6$ and $J_{1a}\!=\!-120\pm20$\,K
for Na$_2$Cu$_2$TeO$_6$, respectively.  For the exchange
between the structural dimers, we find $J_{1b}\!=\!150\pm50$\,K for
Na$_3$Cu$_2$SbO$_6$ and $J_{1b}\!=\!232\pm70$\,K for
Na$_2$Cu$_2$TeO$_6$ ($U_{3d}\!=\!6\mp1$\,eV).  All further exchange integrals
between different chains and layers are smaller than 10\,K, and thus can be
neglected in the minimal model.

Unlike the related compounds featuring edge-shared
chains\cite{FHC_Li2ZrCuO4_DFT_simul,linarite_2012} or Cu$_2$O$_6$
dimers,\cite{cu2po32ch2} Na$_3$Cu$_2$SbO$_6$ and Na$_2$Cu$_2$TeO$_6$ exhibit a
sizable influence of the Coulomb repulsion $U_{3d}$ on the exchange
integrals (see Fig.~\ref{Js_compare}).  However, the variation of $U_{3d}$
within the physically relevant range (Sec.~\ref{S-methods}) does not affect the
FM nature of $J_{1a}$.  Thus, Na$_3$Cu$_2$SbO$_6$ features alternating chains
with the exchange integrals of nearly the same magnitude but different sign (FM
$J_{1a}$ and AFM $J_{1b}$), while for
Na$_2$Cu$_2$TeO$_6$, the AFM exchange between the structural dimers is
dominant. The evaluated exchange integrals are listed in Table~\ref{sbte_Js}.

For an independent computational method, we use hybrid functional (HF)
total energy calculations.  The absence of the double counting problem
and minimal number of free parameters makes HF calculations an appealing
alternative to the DFT+$U$
methods.\cite{2007_HSE03_PBE0_G0W0_Si_InN_ZnO} Here, we employ the HSE06
functional to evaluate the leading couplings $J_{1a}$ and $J_{1b}$ in
both compounds.  In accord with DFT+$U$, we obtain FM $J_{1a}$ and AFM
$J_{1b}$. For \nasb, the resulting exchange integrals are in excellent
agreement with the HTSE estimates (Table~\ref{sbte_Js}). Similar to
DFT+$U$, Na$_2$Cu$_2$TeO$_6$ features a weaker $J_{1a}$ and stronger
$J_{1b}$, thus the $\alpha$ value is substantially reduced.

We are now in position to compare our results with the previous
DFT-based studies.
Derakhshan~\emph{et~al}.~(Ref.~\onlinecite{derakhshan07}) evaluated the
relevant transfer integrals using $N$th-order muffin-tin-orbital
downfolding of the LDA band structure.  Although this computational
method (Ref.~\onlinecite{NMTO}) as well as the code\cite{TB-LMTO-ASA} used for
the calculations differ from our approach, the difference in the
resulting $t_{ij}$ values does not exceed 25\%.\cite{[{
Different DFT codes employ different basis sets, hence the resulting
band structures are not identical.  For an instructive example, see }]
[{}] SCPO_comment}  Hence, the estimated AFM contributions to the
exchanges $J_{1a}$ and $J_{1b}$ generally agree with our values.
However, in contrast to the present study, the authors of
Ref.~\onlinecite{derakhshan07} did not perform DFT+$U$ calculations and
therefore completely disregarded the FM contributions, which are
especially relevant for the short-range coupling $J_{1a}$.  Thus, their
AFM-AFM solution originates from a severe incompleteness of the
computational scheme and the respective mapping onto the spin
Hamiltonian.

In contrast to Ref.~\onlinecite{koo08}, Koo and Whangbo performed
DFT+$U$ calculations using \textsc{vasp}, and recovered FM $J_{1a}$ and
AFM $J_{1b}$, in qualitative agreement with the experiment.  However,
the absolute values of the leading couplings are considerably
overestimated.  We believe that this overestimation stems from the
choice of the on-site Coulomb repulsion parameter $U_d$.  It is
well-known that the parameters of the DFT+$U$ calculations are not
universal,\cite{LSDA+U_Pickett} in particular basis dependent, and
should be carefully chosen based on the nature of the magnetic atom and
the code used.  The $U_d$ range studied in Ref.~\onlinecite{koo08}
(4..7\,eV) is too narrow, and larger $U_d$ is likely required to
reproduce the correct magnetic energy scale in \nasb\ and \nate.  

\begin{table}[tb]
\begin{ruledtabular} \caption{\label{sbte_Js} Leading exchange integrals
$J_{1a}$ and $J_{1b}$ (in K) and the alternation ratio
$\alpha\!\equiv\!J_{1a}/J_{1b}$ for \nasb\ and \nate, evaluated using different
methods.  HTSE and QMC estimates are made based on the experimental data from
the respective reference (first column).  Theoretical estimates, LSDA+$U$,
HSE06, and extended H\"{u}ckel tight-binding (EHTB) are based on calculations
for the experimental crystal structures.}
\begin{tabular}{llrrr}
data source & method & $J_{1a}$ & $J_{1b}$ & $\alpha$\,=\,$J_{1a}$/$J_{1b}$  \\ \hline
\multicolumn{5}{c}{Na$_3$Cu$_2$SbO$_6$} \\ 
\multirow{3}{*}{this study}   & LSDA+$U$  & $-135$ & 150 & $-0.90$ \\
                              & HSE06     & $-205$ & 163 & $-1.26$ \\ 
                              & HTSE      & $-207$ & 171 & $-1.21$ \\ 
                              & QMC       & $-217$ & 174 & $-1.25$ \\ \\
Ref.~\onlinecite{koo08}       & EHTB      & $-165$ & 345 & $-0.48$ \\
Ref.~\onlinecite{miura06}     & HTSE      & $-165$ & 209 & $-0.79$ \\
Ref.~\onlinecite{derakhshan07}& HTSE      &    22  & 169 &   0.13  \\
\hline \multicolumn{5}{c}{Na$_2$Cu$_2$TeO$_6$} \\
\multirow{2}{*}{this study}   & LSDA+$U$  & $-120$ & 232 & $-0.52$ \\
                              & HSE06     & $-165$ & 291 & $-0.57$ \\ \\
Ref.~\onlinecite{koo08}       & EHTB      & $-158$ & 516 & $-0.30$ \\
Ref.~\onlinecite{miura06}     & HTSE      & $-272$ & 215 & $-1.27$ \\
Ref.~\onlinecite{xu05}        & HTSE      &    13  & 127 &   0.1   \\
\end{tabular}
\end{ruledtabular}
\end{table}

\subsection{Influence of chain geometry}
Next, we study the influence of the structural parameters onto the alternation
ratio $\alpha\!=\!J_{1a}/J_{1b}$ for Na$_3$Cu$_2$SbO$_6$ and
Na$_2$Cu$_2$TeO$_6$.  The two compounds differ not only by the nonmagnetic ions
(Sb and Te) located between the structural dimers, but also by details of their
chain geometry.  These subtle differences can have a substantial impact on the
magnetic properties.  In particular, the substitution of Sb by Te and the
corresponding change of the Na content modulates the crystal field. 
Furthermore, the substitution of Sb by Te has a sizable impact on the buckling
of the dimer chains, which is determined by the deviation of O atoms from an
ideal planar arrangement.  Finally, the interatomic distances in the two
compounds are different.  To separate these effects out, we introduce
fictitious compounds containing ideal planar dimer chains (see
Fig.~\ref{F-str}), evaluate their electronic structure, and compare
them with real compounds.

\begin{figure}
\includegraphics[width=8.7cm]{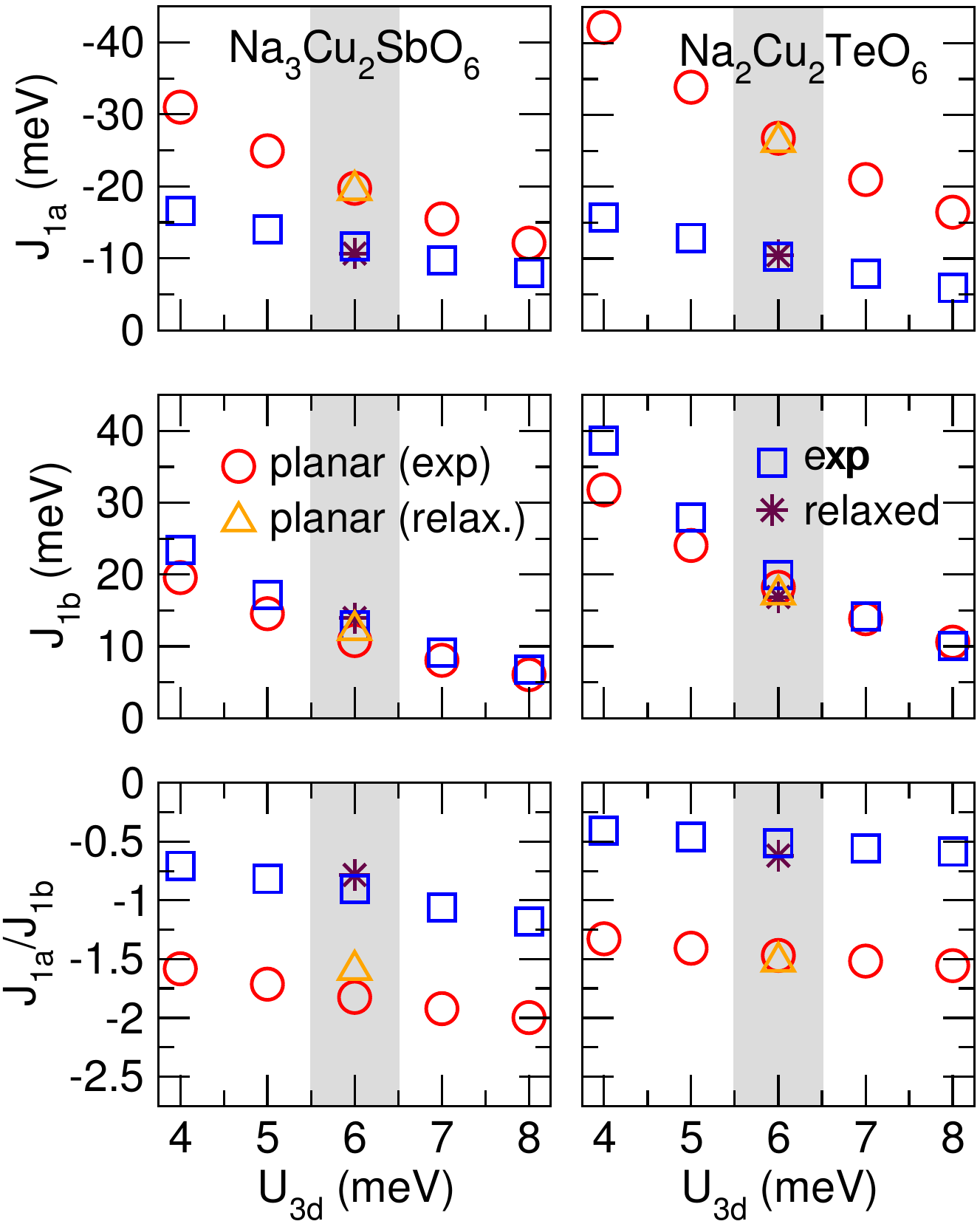}
\caption{\label{Js_compare} 
(Color online) Calculated exchange integrals $J_{1b}$ and $J_{1a}$, as well as
frustration ratios $J_{1a}/J_{1b}\!=\!\alpha$ as a function of the Coulomb
repulsion $U_{3d}$ for different structural models of the two compounds.}
\end{figure}

\begin{figure}
\includegraphics[width=8.7cm]{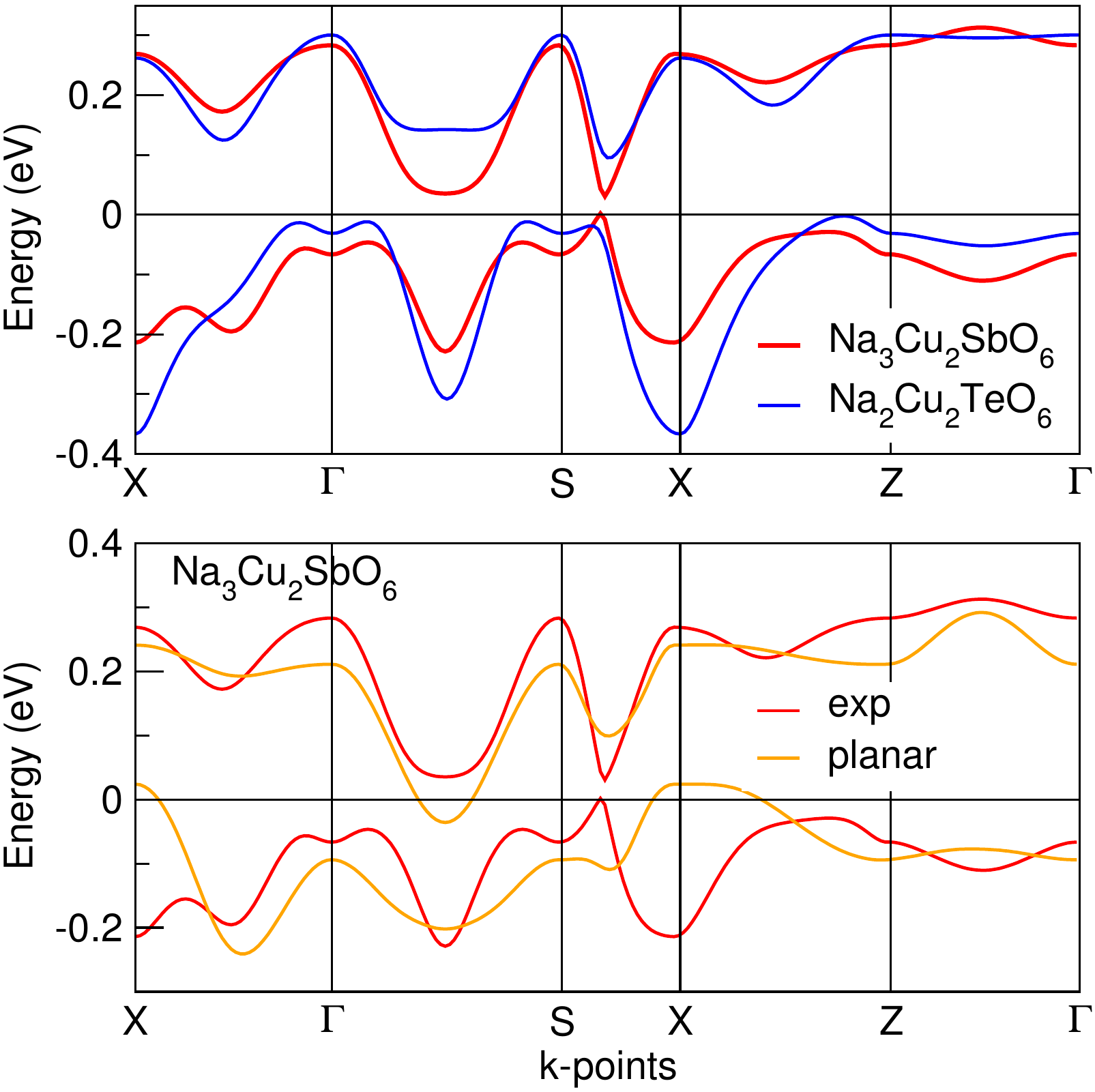}
\caption{\label{vgl_n2n3} 
Top: LDA band structures showing the magnetically active antibonding dp$\sigma^{*}$
states for Na$_3$Cu$_2$SbO$_6$ and Na$_2$Cu$_2$TeO$_6$.  Bottom: comparison of
the LDA band structures calculated for different structural models of
Na$_3$Cu$_2$SbO$_6$.}
\end{figure}

The direct comparison of the antibonding  dp$\sigma^{*}$ bands
for the experimentally observed crystal structures of
Na$_3$Cu$_2$SbO$_6$ and Na$_2$Cu$_2$TeO$_6$ (Fig.~\ref{vgl_n2n3}, upper panel)
reveals that these bands differ mainly by their width. In contrast,
comparing the antibonding  $dp\sigma^{*}$ bands of Na$_3$Cu$_2$SbO$_6$ within the
experimental crystal structure (distorted plaquettes) with the fictitious
crystal structure (planar plaquettes) reveals similar band widths, but
substantially different dispersions (compare X-$\Gamma$ or X-Z in
Fig.~\ref{vgl_n2n3}).  The same trend is also observed for Na$_2$Cu$_2$TeO$_6$.

(i) To separate out the effect of the Sb$\leftrightarrow$Te
substitution, we perform VCA calculations for the same structural model.  In
particular, a certain fraction $x$ of Sb atoms is replaced by Te, with a
concomitant change in the Na content, in order to keep the charge balance. The
band structures calculated for different Te concentrations exhibit similar
dispersions and similar band width, evidencing the minor relevance of the pure
substitutional effect for the magnetic exchange couplings.\cite{suppl}

To estimate the impact of the chain distortion and interatomic distances onto
the magnetism in more detail, we evaluated the magnetic model also for two
fictitious crystal structures of Na$_3$Cu$_2$SbO$_6$ and Na$_2$Cu$_2$TeO$_6$
(featuring planar dimer chains).\cite{suppl} The obtained hopping terms and
exchange integrals are given in Tables~\ref{tab_ts} and \ref{list_Js}.
LSDA+$U$ calculations ($U_{3d}\!=\!6\pm1$\,eV) yield
$J_{1a}$\,=\,$-230\pm50$\,K and $J_{1b}\!=\!126\pm35$\,K for the fictitious
Na$_3$Cu$_2$SbO$_6$ and $J_{1a}\!=\!-312\pm80$\,K and $J_{1b}\!=\!212\pm45$\,K
for the fictitious Na$_2$Cu$_2$TeO$_6$, respectively.  The dependence of the
exchange integrals on the Coulomb repulsion $U_{3d}$ is depicted in
Fig.~\ref{Js_compare}.  Analysis of the resulting exchange couplings suggests
that the two structural parameters act differently: the distortion of the
dimer-chains mainly influences the coupling strength of $J_{1a}$ and the
coupling regime between the dimer-chains ($t_{ic}$ and $t_{il}$), whereas the
interdimer exchange $J_{1b}$ is rather insensitive to this parameter
(Table~\ref{list_Js}), since the respective superexchange path does not involve
O(2) atoms that rule the distortion.

\begin{table}[tb]
\begin{ruledtabular}
\caption{\label{list_Js} DFT estimates for the magnetic exchange integrals in
Na$_3$Cu$_2$SbO$_6$ and Na$_2$Cu$_2$TeO$_6$. The AFM part of the exchange
integral $J^{\text{AFM}}_{1b}$ (in~K) evaluated using the effective one-orbital
model (with $U_{\text{eff}}$\,=\,4\,eV) and total exchange integrals $J_{1b}$
and $J_{1a}$ (in~K) from LSDA+$U$ total energy calculations (using
$U_{3d}\!=\!6.0$\,eV), for the experimental (exp), the LDA-relaxed (relaxed)
and the fictitious (planar) crystal structures.}
\begin{tabular}{lrrr}

\multicolumn{1}{c}{structure} &
\multicolumn{1}{c}{$J_{1a}$} &
\multicolumn{1}{c}{$J_{1b}$} &
\multicolumn{1}{c}{($J^{\text{AFM}}_{1b}$)} \\
\hline
 \multicolumn{4}{c}{Na$_3$Cu$_2$SbO$_6$}\\
 exp     &  $-135$ & 150 & (188) \\
 relaxed& $-125$ &162  & (209)\\
 planar (exp)  & $-230$ & 126 & (165)\\
planar (relax) & $-227$ & 142 & (182)\\
\hline \multicolumn{4}{c}{Na$_2$Cu$_2$TeO$_6$}\\
 exp     &  $-120$ & 232 &(305)\\
relaxed &  $-122$ & 197 & (269)\\
 planar  (exp) &  $-312$ & 212 & (269)\\
planar (relax) & $-305$ & 200 & (255)\\
\end{tabular}
\end{ruledtabular}
\end{table}

(ii) Comparing the total exchange integrals for Na$_3$Cu$_2$SbO$_6$ for the
experimental crystal structure with the planar system discloses an increase of
the NN coupling $J_{1a}$ by nearly a factor of 2, whereas $J_{1b}$ is decreased
by less than 20\%.  This observation is in line with the intuitive picture
derived from geometrical considerations comparing the experimental distorted
crystal structure to the fictitious system containing ideal planar chains
(compare Fig.~\ref{F-str}, lower panel).  Locking the O atoms within the chain
plane directly alters the exchange path of $J_{1a}$  along Cu-O-Cu, by a change
of the Cu-O-Cu bridging angle and the orientation of the magnetically active
orbitals. In contrast, the superexchange path of $J_{1b}$ (Cu-O-O-Cu) is
altered only indirectly by changes of the crystal-field due to the distortion
of the Sb/TeO$_6$ octahedra (compare Fig.~\ref{F-str}, lower panel).

(iii) The modulation of interatomic distances influences $J_{1a}$ and $J_{1b}$ in a
similar way.  The crucial impact of the interatomic distances on $J_{1b}$
manifests itself in the coupling strength of the planar model structures for
Na$_3$Cu$_2$SbO$_6$ and Na$_2$Cu$_2$TeO$_6$ (see Tab.~\ref{list_Js}) with the
corresponding NNN Cu-Cu interdimer distance. The
about 0.1\,\AA\ shorter NNN Cu-Cu distance in the fictitious planar
Na$_2$Cu$_2$TeO$_6$ structure compared to the fictitious planar
Na$_3$Cu$_2$SbO$_6$ increases the coupling strength by about 60\%.
However, comparing the experimental distorted crystal structure with the planar
model structure of Na$_2$Cu$_2$TeO$_6$  the difference in the NNN Cu-Cu
distance is only half as large (about 0.05\,\AA) as between the two planar structures
and result in an about 1/4 smaller increase of $J_{1b}$. Thus, $J_{1b}$ follows
a simple distance relation and scales according to $r^2$.  The same relation
holds for $J_{1a}$ (compare $J_{1a}$ for the two planar structures with the
change of the NN Cu-Cu distance). 

Based on the above considerations, we can conclude that the crucial parameter,
determining the alteration ratio $\alpha\!=\!J_{1a}/J_{1b}$ for
Na$_3$Cu$_2$SbO$_6$ and Na$_2$Cu$_2$TeO$_6$, is the distortion of the chains.
Thus, a directed modification of the chain buckling by the appropriate
substitution of ions should allow to tune the magnetism of these systems.
Furthermore, the chain distortion also influences the interchain coupling
regime. In the experimental structure the long-range exchanges mostly operate
within the magnetic layers (in the $ab$-plane), whereas in the planar system
the coupling between the layers is enhanced ($t_{ij}$'s in Table~\ref{tab_ts}). 

\section{\label{S-spectrum}Energy spectrum} As already mentioned, the FM-AFM
and AFM-AFM solutions correspond to different magnetic GSs. In the former case,
the GS is similar to the Haldane chain and features sizable string order
parameter $O_\text{s}\!=\!0.36$, indicative of a topological
order, while in the latter case the string order is suppressed
($O_\text{s}\!=\!0.16$).\cite{AFHC_theory} It is thus tempting to find an
observable quantity that would be substantially different in the two phases.
Theoretical studies of the $S\!=\!1/2$ AHC model suggest that this requirement
is fulfilled for the momentum position of the spin gap. Indeed, the $Q\!=\!0$
gap is characteristic for AFM-AFM chains, except for the narrow parameter range
$\alpha\!=\!0.79-1.00$, where the gap shifts to small finite
$Q\leq0.02/\pi$.\cite{HC_AHC_Johnston} In contrast, the spin gap in the FM-AFM
chains is located at $Q\!=\!\pi$.\cite{AHC_Hida_FMAF_spectrum} Therefore, by
measuring momentum resolved excitation spectra, the sign of $J_{1a}$ can be
reliably determined.

\begin{figure}[tb]
\includegraphics[width=8.6cm]{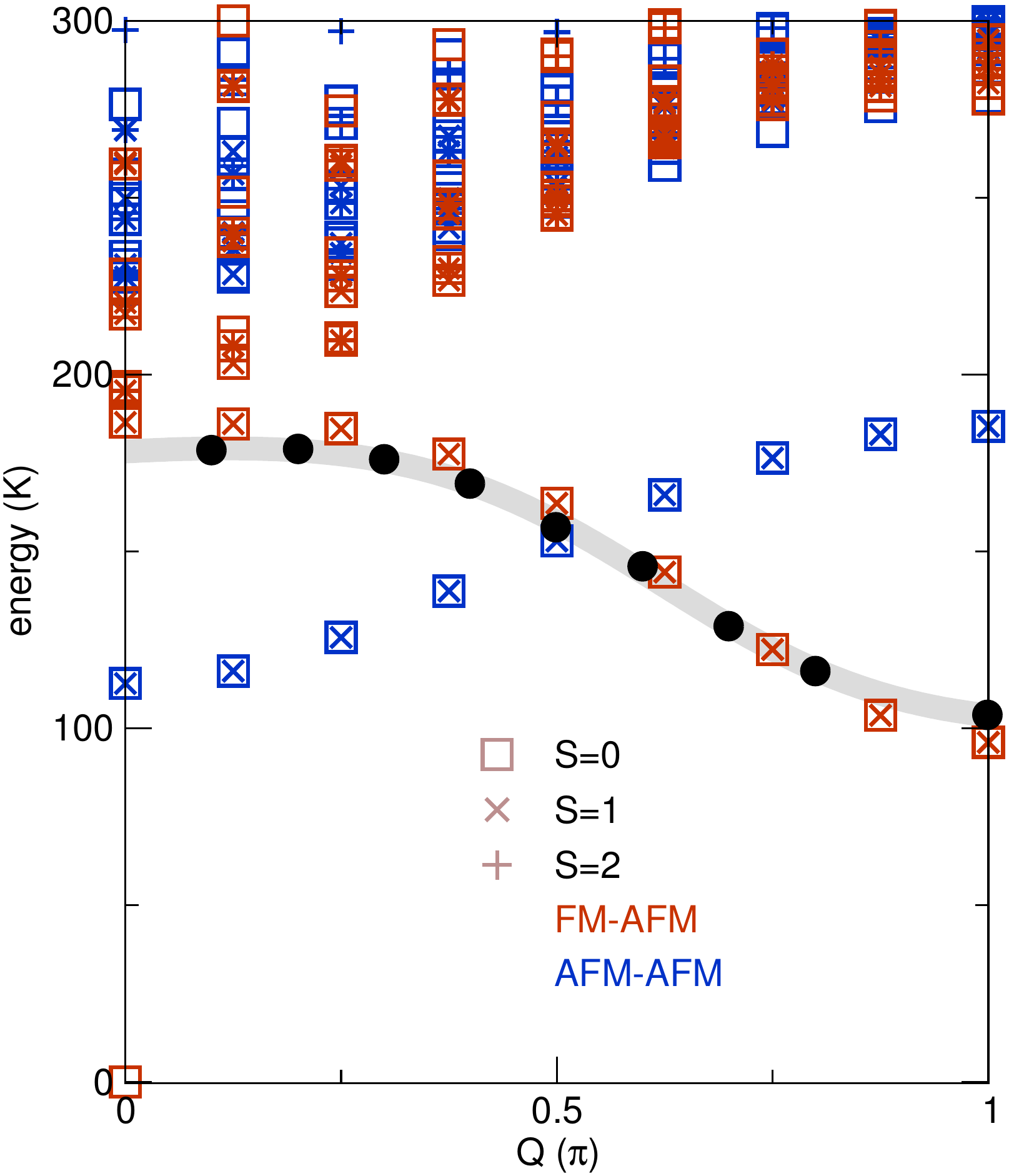}
\caption{\label{F-INS}(Color online) Low-energy excitation spectra for the
alternating Heisenberg chains of $N\!=\!32$ spins, corresponding to the QMC
solutions (FM-AFM and AFM-AFM) from Table~\ref{T-chiT}. Note the difference in
the behavior of the low-lying branch for the two solutions.  Experimental data
from Ref.~\onlinecite{miura08} are shown with filled circles (the gray line is
guide to the eye).}
\end{figure}

To resolve the ambiguity between the FM-AFM and AFM-AFM cases ultimately, we
calculate the low-energy excitations for $\alpha\!=\!-1.25$ as well as
$\alpha\!=\!0.4$ using Lanczos diagonalization of the respective Heisenberg
Hamiltonian. The resulting $E(Q)$ dependence is plotted in Fig.~\ref{F-INS}.
Although the two solutions yield similar estimates for the spin gap, its
$Q$-position is very different: $Q\!=\!\pi$ and $Q\!=\!0$, for the FM-AFM and
AFM-AFM solution, respectively. Another distinct feature of the excitation
spectra is the well-separated branch of lowest-energy excitations
(Fig.~\ref{F-INS}). For the FM-AFM solution, this branch resembles the behavior
of $\cos(Q)$, while the AFM-AFM solution yields a $\sin(Q)$-like behavior.

To compare with the experimental dispersion from Ref.~\onlinecite{miura08}, we scale the two
spectra using the values of the exchange couplings from our QMC fits to the
magnetic susceptibility (Table~\ref{T-chiT}, last row). This way, we find that
the FM-AFM solution agrees very well with the experimental data (Fig.~\ref{F-INS}), while
the AFM-AFM solution can be safely ruled out. 

Fig.~\ref{F-INS} also provides an answer to an intriguing question, why both
$\alpha\!=\!-1.25$ and $\alpha\!=\!0.4$ provide good fits to the susceptibility
data.  At finite temperature, magnetic susceptibility reflects the
thermal-averaged magnetic spectrum integrated over the whole momentum space.
Thus, at low temperatures, $\chi(T)$ is largely affected by the value of the
spin gap, but is insensitive to its $Q$ position.  Since the values of the spin
gap for the two solutions are very similar (around 100\,K), the similarity of
the low-temperature $\chi(T)$ is also not surprising. Moreover, the shape of
the low-energy branch is similar (but reflected around $Q\!=\!\pi/2$), thus the
$Q$-integrated spectrum is nearly the same in both cases.  Only at elevated
temperatures, the contribution of high-lying states gives rise to the
difference in $\chi(T)$.  This is in excellent agreement with the enhanced
deviation of the AFM-AFM solution at high temperatures (Fig.~\ref{F-chiT},
inset).

\section{\label{S-sum}Summary}
Since the first report on the magnetism of the low-dimensional
$S$\,=\,1/2 systems \nasb\ and \nate, their spin models were
controversially debated in the literature.  The main conundrum was the
sign of the exchange coupling $J_{\text{1a}}$ operating within the
structural Cu$_2$O$_6$ dimers.  To resolve the conflicting reports, we
applied a series of different computational methods, including density
functional theory (DFT) band structure, virtual crystal approximation,
DFT+$U$, and hybrid functional calculations, as well as high-temperature
series expansions, quantum Monte Carlo simulations, and exact
diagonalization.

Our calculations evidence that the magnetism of both compounds can be
described by the alternating Heisenberg chain model with two relevant
couplings: ferromagnetic $J_{\text{1a}}$ within the structural dimers,
and antiferromagnetic $J_{\text{1b}}$ between the dimers.  The
alternation parameter $\alpha\!=\!J_{\text{1a}}/J_{\text{1b}}$ amounts
to about $-1.25$ and $-0.55$  in \nasb\ and \nate, respectively. This
parameter regime corresponds to the Haldane phase, characterized by the
gapped excitation spectrum and a topological string order.

Using extensive calculations for different structural models, we find
that the physically relevant ratio $\alpha\!=\!J_{1a}/J_{1b}$ is
primarily ruled by the distortion of the structural chains, while the
Sb$\leftrightarrow$Te substitution and the change in the Cu--Cu distance
play a minor role.  A comparison of the simulated dispersion
$E(Q)$ with the experimental inelastic neutron scattering data
(Ref.~\onlinecite{miura08}) yields an unequivocal evidence for the FM
nature of $J_{\text{1a}}$ in \nasb.  These spectra facilitate the
understanding of the similarity between the magnetic susceptibility
curves for mutually exclusive solutions that involve ferromagnetic and
antiferromagnetic $J_{\text{1a}}$.

It is important to note that the problem of ambiguous solutions appears
in the empirical modeling, only.  In contrast, the microscopic modeling
based on DFT calculations readily yields a quasi-one-dimensional model
with the ferromagnetic $J_{\text{1a}}$.  This clearly indicates that
present-day DFT calculations are a reliable tool to disclose even rather
complicated cases and deliver a reliable microscopic magnetic model.
Since the correctness of the magnetic model is of crucial importance for
its refinement and extension, DFT calculations should be an
indispensable ingredient of real-material studies.

\section*{Acknowledgment}
We are grateful to Professor M. Sato for explaining the details of his
previous experimental work on \nasb\ and \nate.  Fruitful discussions
with A.~A.~Tsirlin are kindly acknowledged.  OJ was partially supported
by the European Union through the European Social Fund (Mobilitas Grant
no.  MJD447). JR was supported by the DFG through the project RI
615/16-3.


\begin{thebibliography}{53}%
\makeatletter
\providecommand \@ifxundefined [1]{%
 \@ifx{#1\undefined}
}%
\providecommand \@ifnum [1]{%
 \ifnum #1\expandafter \@firstoftwo
 \else \expandafter \@secondoftwo
 \fi
}%
\providecommand \@ifx [1]{%
 \ifx #1\expandafter \@firstoftwo
 \else \expandafter \@secondoftwo
 \fi
}%
\providecommand \natexlab [1]{#1}%
\providecommand \enquote  [1]{``#1''}%
\providecommand \bibnamefont  [1]{#1}%
\providecommand \bibfnamefont [1]{#1}%
\providecommand \citenamefont [1]{#1}%
\providecommand \href@noop [0]{\@secondoftwo}%
\providecommand \href [0]{\begingroup \@sanitize@url \@href}%
\providecommand \@href[1]{\@@startlink{#1}\@@href}%
\providecommand \@@href[1]{\endgroup#1\@@endlink}%
\providecommand \@sanitize@url [0]{\catcode `\\12\catcode `\$12\catcode
  `\&12\catcode `\#12\catcode `\^12\catcode `\_12\catcode `\%12\relax}%
\providecommand \@@startlink[1]{}%
\providecommand \@@endlink[0]{}%
\providecommand \url  [0]{\begingroup\@sanitize@url \@url }%
\providecommand \@url [1]{\endgroup\@href {#1}{\urlprefix }}%
\providecommand \urlprefix  [0]{URL }%
\providecommand \Eprint [0]{\href }%
\providecommand \doibase [0]{http://dx.doi.org/}%
\providecommand \selectlanguage [0]{\@gobble}%
\providecommand \bibinfo  [0]{\@secondoftwo}%
\providecommand \bibfield  [0]{\@secondoftwo}%
\providecommand \translation [1]{[#1]}%
\providecommand \BibitemOpen [0]{}%
\providecommand \bibitemStop [0]{}%
\providecommand \bibitemNoStop [0]{.\EOS\space}%
\providecommand \EOS [0]{\spacefactor3000\relax}%
\providecommand \BibitemShut  [1]{\csname bibitem#1\endcsname}%
\let\auto@bib@innerbib\@empty
\bibitem [{\citenamefont {Lee}(2008)}]{lee08}%
  \BibitemOpen
  \bibfield  {author} {\bibinfo {author} {\bibfnamefont {P.~A.}\ \bibnamefont
  {Lee}},\ }\href {\doibase 10.1088/0034-4885/71/1/012501} {\bibfield
  {journal} {\bibinfo  {journal} {Rep. Prog. Phys.}\ }\textbf {\bibinfo
  {volume} {71}},\ \bibinfo {pages} {012501} (\bibinfo {year} {2008})},\
  \Eprint {http://arxiv.org/abs/arXiv:0708.2115} {arXiv:0708.2115} \BibitemShut
  {NoStop}%
\bibitem [{\citenamefont {Balents}(2010)}]{balents10}%
  \BibitemOpen
  \bibfield  {author} {\bibinfo {author} {\bibfnamefont {L.}~\bibnamefont
  {Balents}},\ }\href {\doibase 10.1038/nature08917} {\bibfield  {journal}
  {\bibinfo  {journal} {Nature (London)}\ }\textbf {\bibinfo {volume} {464}},\
  \bibinfo {pages} {199} (\bibinfo {year} {2010})}\BibitemShut {NoStop}%
\bibitem [{\citenamefont {Helton}\ \emph {et~al.}(2007)\citenamefont {Helton},
  \citenamefont {Matan}, \citenamefont {Shores}, \citenamefont {Nytko},
  \citenamefont {Bartlett}, \citenamefont {Yoshida}, \citenamefont {Takano},
  \citenamefont {Suslov}, \citenamefont {Qiu}, \citenamefont {Chung},
  \citenamefont {Nocera},\ and\ \citenamefont {Lee}}]{herb}%
  \BibitemOpen
  \bibfield  {author} {\bibinfo {author} {\bibfnamefont {J.~S.}\ \bibnamefont
  {Helton}}, \bibinfo {author} {\bibfnamefont {K.}~\bibnamefont {Matan}},
  \bibinfo {author} {\bibfnamefont {M.~P.}\ \bibnamefont {Shores}}, \bibinfo
  {author} {\bibfnamefont {E.~A.}\ \bibnamefont {Nytko}}, \bibinfo {author}
  {\bibfnamefont {B.~M.}\ \bibnamefont {Bartlett}}, \bibinfo {author}
  {\bibfnamefont {Y.}~\bibnamefont {Yoshida}}, \bibinfo {author} {\bibfnamefont
  {Y.}~\bibnamefont {Takano}}, \bibinfo {author} {\bibfnamefont
  {A.}~\bibnamefont {Suslov}}, \bibinfo {author} {\bibfnamefont
  {Y.}~\bibnamefont {Qiu}}, \bibinfo {author} {\bibfnamefont {J.-H.}\
  \bibnamefont {Chung}}, \bibinfo {author} {\bibfnamefont {D.~G.}\ \bibnamefont
  {Nocera}}, \ and\ \bibinfo {author} {\bibfnamefont {Y.~S.}\ \bibnamefont
  {Lee}},\ }\href {\doibase 10.1103/PhysRevLett.98.107204} {\bibfield
  {journal} {\bibinfo  {journal} {Phys. Rev. Lett.}\ }\textbf {\bibinfo
  {volume} {98}},\ \bibinfo {pages} {107204} (\bibinfo {year} {2007})},\
  \Eprint {http://arxiv.org/abs/cond-mat/0610539} {cond-mat/0610539}
  \BibitemShut {NoStop}%
\bibitem [{\citenamefont {Nakatsuji}\ \emph {et~al.}(2012)\citenamefont
  {Nakatsuji}, \citenamefont {Kuga}, \citenamefont {Kimura}, \citenamefont
  {Satake}, \citenamefont {Katayama}, \citenamefont {Nishibori}, \citenamefont
  {Sawa}, \citenamefont {Ishii}, \citenamefont {Hagiwara}, \citenamefont
  {Bridges}, \citenamefont {Ito}, \citenamefont {Higemoto}, \citenamefont
  {Karaki}, \citenamefont {Halim}, \citenamefont {Nugroho}, \citenamefont
  {Rodriguez-Rivera}, \citenamefont {Green},\ and\ \citenamefont
  {Broholm}}]{na3cusb2o9}%
  \BibitemOpen
  \bibfield  {author} {\bibinfo {author} {\bibfnamefont {S.}~\bibnamefont
  {Nakatsuji}}, \bibinfo {author} {\bibfnamefont {K.}~\bibnamefont {Kuga}},
  \bibinfo {author} {\bibfnamefont {K.}~\bibnamefont {Kimura}}, \bibinfo
  {author} {\bibfnamefont {R.}~\bibnamefont {Satake}}, \bibinfo {author}
  {\bibfnamefont {N.}~\bibnamefont {Katayama}}, \bibinfo {author}
  {\bibfnamefont {E.}~\bibnamefont {Nishibori}}, \bibinfo {author}
  {\bibfnamefont {H.}~\bibnamefont {Sawa}}, \bibinfo {author} {\bibfnamefont
  {R.}~\bibnamefont {Ishii}}, \bibinfo {author} {\bibfnamefont
  {M.}~\bibnamefont {Hagiwara}}, \bibinfo {author} {\bibfnamefont
  {F.}~\bibnamefont {Bridges}}, \bibinfo {author} {\bibfnamefont {T.~U.}\
  \bibnamefont {Ito}}, \bibinfo {author} {\bibfnamefont {W.}~\bibnamefont
  {Higemoto}}, \bibinfo {author} {\bibfnamefont {Y.}~\bibnamefont {Karaki}},
  \bibinfo {author} {\bibfnamefont {M.}~\bibnamefont {Halim}}, \bibinfo
  {author} {\bibfnamefont {A.~A.}\ \bibnamefont {Nugroho}}, \bibinfo {author}
  {\bibfnamefont {J.~A.}\ \bibnamefont {Rodriguez-Rivera}}, \bibinfo {author}
  {\bibfnamefont {M.~A.}\ \bibnamefont {Green}}, \ and\ \bibinfo {author}
  {\bibfnamefont {C.}~\bibnamefont {Broholm}},\ }\href {\doibase
  10.1126/science.1212154} {\bibfield  {journal} {\bibinfo  {journal}
  {Science}\ }\textbf {\bibinfo {volume} {336}},\ \bibinfo {pages} {559}
  (\bibinfo {year} {2012})}\BibitemShut {NoStop}%
\bibitem [{\citenamefont {Isobe}\ and\ \citenamefont
  {Ueda}(1996)}]{csv2o5_isobe96}%
  \BibitemOpen
  \bibfield  {author} {\bibinfo {author} {\bibfnamefont {M.}~\bibnamefont
  {Isobe}}\ and\ \bibinfo {author} {\bibfnamefont {Y.}~\bibnamefont {Ueda}},\
  }\href {\doibase 10.1143/JPSJ.65.3142} {\bibfield  {journal} {\bibinfo
  {journal} {J. Phys. Soc. Jpn.}\ }\textbf {\bibinfo {volume} {65}},\ \bibinfo
  {pages} {3142} (\bibinfo {year} {1996})}\BibitemShut {NoStop}%
\bibitem [{\citenamefont {Valent{\'{i}}}\ and\ \citenamefont
  {Saha-Dasgupta}(2002)}]{csv2o5_valenti02}%
  \BibitemOpen
  \bibfield  {author} {\bibinfo {author} {\bibfnamefont {R.}~\bibnamefont
  {Valent{\'{i}}}}\ and\ \bibinfo {author} {\bibfnamefont {T.}~\bibnamefont
  {Saha-Dasgupta}},\ }\href {\doibase 10.1103/PhysRevB.65.144445} {\bibfield
  {journal} {\bibinfo  {journal} {Phys. Rev. B}\ }\textbf {\bibinfo {volume}
  {65}},\ \bibinfo {pages} {144445} (\bibinfo {year} {2002})},\ \Eprint
  {http://arxiv.org/abs/cond-mat/0110311} {cond-mat/0110311} \BibitemShut
  {NoStop}%
\bibitem [{\citenamefont {Sa{\'{u}}l}\ and\ \citenamefont
  {Radtke}(2011)}]{csv2o5_saul11}%
  \BibitemOpen
  \bibfield  {author} {\bibinfo {author} {\bibfnamefont {A.}~\bibnamefont
  {Sa{\'{u}}l}}\ and\ \bibinfo {author} {\bibfnamefont {G.}~\bibnamefont
  {Radtke}},\ }\href {\doibase 10.1103/PhysRevLett.106.177203} {\bibfield
  {journal} {\bibinfo  {journal} {Phys. Rev. Lett.}\ }\textbf {\bibinfo
  {volume} {106}},\ \bibinfo {pages} {177203} (\bibinfo {year}
  {2011})}\BibitemShut {NoStop}%
\bibitem [{\citenamefont {Deisenhofer}\ \emph {et~al.}(2006)\citenamefont
  {Deisenhofer}, \citenamefont {Eremina}, \citenamefont {Pimenov},
  \citenamefont {Gavrilova}, \citenamefont {Berger}, \citenamefont {Johnsson},
  \citenamefont {Lemmens}, \citenamefont {{Krug von Nidda}}, \citenamefont
  {Loidl}, \citenamefont {Lee},\ and\ \citenamefont
  {Whangbo}}]{cuteO_deisenhofer06}%
  \BibitemOpen
  \bibfield  {author} {\bibinfo {author} {\bibfnamefont {J.}~\bibnamefont
  {Deisenhofer}}, \bibinfo {author} {\bibfnamefont {R.~M.}\ \bibnamefont
  {Eremina}}, \bibinfo {author} {\bibfnamefont {A.}~\bibnamefont {Pimenov}},
  \bibinfo {author} {\bibfnamefont {T.}~\bibnamefont {Gavrilova}}, \bibinfo
  {author} {\bibfnamefont {H.}~\bibnamefont {Berger}}, \bibinfo {author}
  {\bibfnamefont {M.}~\bibnamefont {Johnsson}}, \bibinfo {author}
  {\bibfnamefont {P.}~\bibnamefont {Lemmens}}, \bibinfo {author} {\bibfnamefont
  {H.-A.}\ \bibnamefont {{Krug von Nidda}}}, \bibinfo {author} {\bibfnamefont
  {A.}~\bibnamefont {Loidl}}, \bibinfo {author} {\bibfnamefont {K.-S.}\
  \bibnamefont {Lee}}, \ and\ \bibinfo {author} {\bibfnamefont {M.-H.}\
  \bibnamefont {Whangbo}},\ }\href {\doibase 10.1103/PhysRevB.74.174421}
  {\bibfield  {journal} {\bibinfo  {journal} {Phys. Rev. B}\ }\textbf {\bibinfo
  {volume} {74}},\ \bibinfo {pages} {174421} (\bibinfo {year} {2006})},\
  \Eprint {http://arxiv.org/abs/cond-mat/0610458} {cond-mat/0610458}
  \BibitemShut {NoStop}%
\bibitem [{\citenamefont {Das}\ \emph {et~al.}(2008)\citenamefont {Das},
  \citenamefont {Saha-Dasgupta}, \citenamefont {Gros},\ and\ \citenamefont
  {Valent{\'{i}}}}]{cuteo_das08}%
  \BibitemOpen
  \bibfield  {author} {\bibinfo {author} {\bibfnamefont {H.}~\bibnamefont
  {Das}}, \bibinfo {author} {\bibfnamefont {T.}~\bibnamefont {Saha-Dasgupta}},
  \bibinfo {author} {\bibfnamefont {C.}~\bibnamefont {Gros}}, \ and\ \bibinfo
  {author} {\bibfnamefont {R.}~\bibnamefont {Valent{\'{i}}}},\ }\href {\doibase
  10.1103/PhysRevB.77.224437} {\bibfield  {journal} {\bibinfo  {journal} {Phys.
  Rev. B}\ }\textbf {\bibinfo {volume} {77}},\ \bibinfo {pages} {224437}
  (\bibinfo {year} {2008})},\ \Eprint {http://arxiv.org/abs/cond-mat/0703675}
  {cond-mat/0703675} \BibitemShut {NoStop}%
\bibitem [{\citenamefont {Ushakov}\ and\ \citenamefont
  {Streltsov}(2009)}]{cuteo_ushakov09}%
  \BibitemOpen
  \bibfield  {author} {\bibinfo {author} {\bibfnamefont {A.~V.}\ \bibnamefont
  {Ushakov}}\ and\ \bibinfo {author} {\bibfnamefont {S.~V.}\ \bibnamefont
  {Streltsov}},\ }\href {\doibase 10.1088/0953-8984/21/30/305501} {\bibfield
  {journal} {\bibinfo  {journal} {J. Phys.: Condens. Matter}\ }\textbf
  {\bibinfo {volume} {21}},\ \bibinfo {pages} {305501} (\bibinfo {year}
  {2009})}\BibitemShut {NoStop}%
\bibitem [{\citenamefont {Sasago}\ \emph {et~al.}(1995)\citenamefont {Sasago},
  \citenamefont {Hase}, \citenamefont {Uchinokura}, \citenamefont {Tokunaga},\
  and\ \citenamefont {Miura}}]{cacuge2O6_sasago95}%
  \BibitemOpen
  \bibfield  {author} {\bibinfo {author} {\bibfnamefont {Y.}~\bibnamefont
  {Sasago}}, \bibinfo {author} {\bibfnamefont {M.}~\bibnamefont {Hase}},
  \bibinfo {author} {\bibfnamefont {K.}~\bibnamefont {Uchinokura}}, \bibinfo
  {author} {\bibfnamefont {M.}~\bibnamefont {Tokunaga}}, \ and\ \bibinfo
  {author} {\bibfnamefont {N.}~\bibnamefont {Miura}},\ }\href {\doibase
  10.1103/PhysRevB.52.3533} {\bibfield  {journal} {\bibinfo  {journal} {Phys.
  Rev. B}\ }\textbf {\bibinfo {volume} {52}},\ \bibinfo {pages} {3533}
  (\bibinfo {year} {1995})}\BibitemShut {NoStop}%
\bibitem [{\citenamefont {Valent{\'{i}}}\ \emph {et~al.}(2002)\citenamefont
  {Valent{\'{i}}}, \citenamefont {Saha-Dasgupta},\ and\ \citenamefont
  {Gros}}]{cacuge2o6_valenti02}%
  \BibitemOpen
  \bibfield  {author} {\bibinfo {author} {\bibfnamefont {R.}~\bibnamefont
  {Valent{\'{i}}}}, \bibinfo {author} {\bibfnamefont {T.}~\bibnamefont
  {Saha-Dasgupta}}, \ and\ \bibinfo {author} {\bibfnamefont {C.}~\bibnamefont
  {Gros}},\ }\href {\doibase 10.1103/PhysRevB.66.054426} {\bibfield  {journal}
  {\bibinfo  {journal} {Phys. Rev. B}\ }\textbf {\bibinfo {volume} {66}},\
  \bibinfo {pages} {054426} (\bibinfo {year} {2002})},\ \Eprint
  {http://arxiv.org/abs/cond-mat/0202310} {cond-mat/0202310} \BibitemShut
  {NoStop}%
\bibitem [{\citenamefont {Schmitt}\ \emph {et~al.}(2010)\citenamefont
  {Schmitt}, \citenamefont {Gippius}, \citenamefont {Okhotnikov}, \citenamefont
  {Schnelle}, \citenamefont {Koch}, \citenamefont {Janson}, \citenamefont
  {Liu}, \citenamefont {Huang}, \citenamefont {Skourski}, \citenamefont
  {Weickert}, \citenamefont {Baenitz},\ and\ \citenamefont
  {Rosner}}]{cu2po32ch2}%
  \BibitemOpen
  \bibfield  {author} {\bibinfo {author} {\bibfnamefont {M.}~\bibnamefont
  {Schmitt}}, \bibinfo {author} {\bibfnamefont {A.~A.}\ \bibnamefont
  {Gippius}}, \bibinfo {author} {\bibfnamefont {K.~S.}\ \bibnamefont
  {Okhotnikov}}, \bibinfo {author} {\bibfnamefont {W.}~\bibnamefont
  {Schnelle}}, \bibinfo {author} {\bibfnamefont {K.}~\bibnamefont {Koch}},
  \bibinfo {author} {\bibfnamefont {O.}~\bibnamefont {Janson}}, \bibinfo
  {author} {\bibfnamefont {W.}~\bibnamefont {Liu}}, \bibinfo {author}
  {\bibfnamefont {Y.-H.}\ \bibnamefont {Huang}}, \bibinfo {author}
  {\bibfnamefont {Y.}~\bibnamefont {Skourski}}, \bibinfo {author}
  {\bibfnamefont {F.}~\bibnamefont {Weickert}}, \bibinfo {author}
  {\bibfnamefont {M.}~\bibnamefont {Baenitz}}, \ and\ \bibinfo {author}
  {\bibfnamefont {H.}~\bibnamefont {Rosner}},\ }\href {\doibase
  10.1103/PhysRevB.81.104416} {\bibfield  {journal} {\bibinfo  {journal} {Phys.
  Rev. B}\ }\textbf {\bibinfo {volume} {81}},\ \bibinfo {pages} {104416}
  (\bibinfo {year} {2010})}\BibitemShut {NoStop}%
\bibitem [{\citenamefont {Mazurenko}\ \emph {et~al.}(2014)\citenamefont
  {Mazurenko}, \citenamefont {Valentyuk}, \citenamefont {Stern},\ and\
  \citenamefont {Tsirlin}}]{mazurenko2013}%
  \BibitemOpen
  \bibfield  {author} {\bibinfo {author} {\bibfnamefont {V.~V.}\ \bibnamefont
  {Mazurenko}}, \bibinfo {author} {\bibfnamefont {M.~V.}\ \bibnamefont
  {Valentyuk}}, \bibinfo {author} {\bibfnamefont {R.}~\bibnamefont {Stern}}, \
  and\ \bibinfo {author} {\bibfnamefont {A.~A.}\ \bibnamefont {Tsirlin}},\
  }\href {\doibase 10.1103/PhysRevLett.112.107202} {} (\bibinfo {year}
  {2014}),\ \Eprint {http://arxiv.org/abs/arXiv:1309.6762} {arXiv:1309.6762}
  \BibitemShut {NoStop}%
\bibitem [{\citenamefont {Jaime}\ \emph {et~al.}(2004)\citenamefont {Jaime},
  \citenamefont {Correa}, \citenamefont {Harrison}, \citenamefont {Batista},
  \citenamefont {Kawashima}, \citenamefont {Kazuma}, \citenamefont {Jorge},
  \citenamefont {Stern}, \citenamefont {Heinmaa}, \citenamefont {Zvyagin},
  \citenamefont {Sasago},\ and\ \citenamefont
  {Uchinokura}}]{bacusi2O6_jaime06}%
  \BibitemOpen
  \bibfield  {author} {\bibinfo {author} {\bibfnamefont {M.}~\bibnamefont
  {Jaime}}, \bibinfo {author} {\bibfnamefont {V.~F.}\ \bibnamefont {Correa}},
  \bibinfo {author} {\bibfnamefont {N.}~\bibnamefont {Harrison}}, \bibinfo
  {author} {\bibfnamefont {C.~D.}\ \bibnamefont {Batista}}, \bibinfo {author}
  {\bibfnamefont {N.}~\bibnamefont {Kawashima}}, \bibinfo {author}
  {\bibfnamefont {Y.}~\bibnamefont {Kazuma}}, \bibinfo {author} {\bibfnamefont
  {G.~A.}\ \bibnamefont {Jorge}}, \bibinfo {author} {\bibfnamefont
  {R.}~\bibnamefont {Stern}}, \bibinfo {author} {\bibfnamefont
  {I.}~\bibnamefont {Heinmaa}}, \bibinfo {author} {\bibfnamefont {S.~A.}\
  \bibnamefont {Zvyagin}}, \bibinfo {author} {\bibfnamefont {Y.}~\bibnamefont
  {Sasago}}, \ and\ \bibinfo {author} {\bibfnamefont {K.}~\bibnamefont
  {Uchinokura}},\ }\href {\doibase 10.1103/PhysRevLett.93.087203} {\bibfield
  {journal} {\bibinfo  {journal} {Phys. Rev. Lett.}\ }\textbf {\bibinfo
  {volume} {93}},\ \bibinfo {pages} {087203} (\bibinfo {year} {2004})},\
  \Eprint {http://arxiv.org/abs/cond-mat/0404324} {cond-mat/0404324}
  \BibitemShut {NoStop}%
\bibitem [{\citenamefont {Sebastian}\ \emph {et~al.}(2006)\citenamefont
  {Sebastian}, \citenamefont {Harrison}, \citenamefont {Batista}, \citenamefont
  {Balicas}, \citenamefont {Jaime}, \citenamefont {Sharma}, \citenamefont
  {Kawashima},\ and\ \citenamefont {Fisher}}]{bacusi2o6_sebastian06}%
  \BibitemOpen
  \bibfield  {author} {\bibinfo {author} {\bibfnamefont {S.~E.}\ \bibnamefont
  {Sebastian}}, \bibinfo {author} {\bibfnamefont {N.}~\bibnamefont {Harrison}},
  \bibinfo {author} {\bibfnamefont {C.~D.}\ \bibnamefont {Batista}}, \bibinfo
  {author} {\bibfnamefont {L.}~\bibnamefont {Balicas}}, \bibinfo {author}
  {\bibfnamefont {M.}~\bibnamefont {Jaime}}, \bibinfo {author} {\bibfnamefont
  {P.~A.}\ \bibnamefont {Sharma}}, \bibinfo {author} {\bibfnamefont
  {N.}~\bibnamefont {Kawashima}}, \ and\ \bibinfo {author} {\bibfnamefont
  {I.~R.}\ \bibnamefont {Fisher}},\ }\href {\doibase 10.1038/nature04732}
  {\bibfield  {journal} {\bibinfo  {journal} {Nature (London)}\ }\textbf
  {\bibinfo {volume} {441}},\ \bibinfo {pages} {617} (\bibinfo {year}
  {2006})},\ \Eprint {http://arxiv.org/abs/cond-mat/0606042} {cond-mat/0606042}
  \BibitemShut {NoStop}%
\bibitem [{\citenamefont {Kr{\"{a}}mer}\ \emph {et~al.}(2007)\citenamefont
  {Kr{\"{a}}mer}, \citenamefont {Stern}, \citenamefont {M.}, \citenamefont
  {Berthier}, \citenamefont {Kimura},\ and\ \citenamefont
  {Fisher}}]{bacusi2o6_kraemer07}%
  \BibitemOpen
  \bibfield  {author} {\bibinfo {author} {\bibfnamefont {S.}~\bibnamefont
  {Kr{\"{a}}mer}}, \bibinfo {author} {\bibfnamefont {R.}~\bibnamefont {Stern}},
  \bibinfo {author} {\bibfnamefont {H.}~\bibnamefont {M.}}, \bibinfo {author}
  {\bibfnamefont {C.}~\bibnamefont {Berthier}}, \bibinfo {author}
  {\bibfnamefont {T.}~\bibnamefont {Kimura}}, \ and\ \bibinfo {author}
  {\bibfnamefont {I.~R.}\ \bibnamefont {Fisher}},\ }\href {\doibase
  10.1103/PhysRevB.76.100406} {\bibfield  {journal} {\bibinfo  {journal} {Phys.
  Rev. B}\ }\textbf {\bibinfo {volume} {76}},\ \bibinfo {pages} {100406}
  (\bibinfo {year} {2007})},\ \Eprint {http://arxiv.org/abs/arXiv:0704.0888}
  {arXiv:0704.0888} \BibitemShut {NoStop}%
\bibitem [{\citenamefont {Takigawa}\ \emph {et~al.}(2010)\citenamefont
  {Takigawa}, \citenamefont {Waki}, \citenamefont {Horvati{\'{c}}},\ and\
  \citenamefont {Berthier}}]{SCBO_NMR_review}%
  \BibitemOpen
  \bibfield  {author} {\bibinfo {author} {\bibfnamefont {M.}~\bibnamefont
  {Takigawa}}, \bibinfo {author} {\bibfnamefont {T.}~\bibnamefont {Waki}},
  \bibinfo {author} {\bibfnamefont {M.}~\bibnamefont {Horvati{\'{c}}}}, \ and\
  \bibinfo {author} {\bibfnamefont {C.}~\bibnamefont {Berthier}},\ }\href
  {\doibase 10.1143/JPSJ.79.011005} {\bibfield  {journal} {\bibinfo  {journal}
  {J. Phys. Soc. Jpn.}\ }\textbf {\bibinfo {volume} {79}},\ \bibinfo {pages}
  {011005} (\bibinfo {year} {2010})}\BibitemShut {NoStop}%
\bibitem [{\citenamefont {Goodenough}(1955)}]{gka_1}%
  \BibitemOpen
  \bibfield  {author} {\bibinfo {author} {\bibfnamefont {J.~B.}\ \bibnamefont
  {Goodenough}},\ }\href {\doibase 10.1103/PhysRev.100.564} {\bibfield
  {journal} {\bibinfo  {journal} {Phys. Rev.}\ }\textbf {\bibinfo {volume}
  {100}},\ \bibinfo {pages} {564} (\bibinfo {year} {1955})}\BibitemShut
  {NoStop}%
\bibitem [{\citenamefont {Kanamori}(1959)}]{gka_2}%
  \BibitemOpen
  \bibfield  {author} {\bibinfo {author} {\bibfnamefont {J.}~\bibnamefont
  {Kanamori}},\ }\href {\doibase 10.1016/0022-3697(59)90061-7} {\bibfield
  {journal} {\bibinfo  {journal} {J. Phys. Chem. Solids}\ }\textbf {\bibinfo
  {volume} {10}},\ \bibinfo {pages} {87} (\bibinfo {year} {1959})}\BibitemShut
  {NoStop}%
\bibitem [{\citenamefont {Pickett}(1989)}]{HTSC_Pickett}%
  \BibitemOpen
  \bibfield  {author} {\bibinfo {author} {\bibfnamefont {W.~E.}\ \bibnamefont
  {Pickett}},\ }\href {\doibase 10.1103/RevModPhys.61.433} {\bibfield
  {journal} {\bibinfo  {journal} {Rev. Mod. Phys.}\ }\textbf {\bibinfo {volume}
  {61}},\ \bibinfo {pages} {433} (\bibinfo {year} {1989})}\BibitemShut
  {NoStop}%
\bibitem [{\citenamefont {Miura}\ \emph {et~al.}(2006)\citenamefont {Miura},
  \citenamefont {Hirai}, \citenamefont {Kobayashi},\ and\ \citenamefont
  {Sato}}]{miura06}%
  \BibitemOpen
  \bibfield  {author} {\bibinfo {author} {\bibfnamefont {Y.}~\bibnamefont
  {Miura}}, \bibinfo {author} {\bibfnamefont {R.}~\bibnamefont {Hirai}},
  \bibinfo {author} {\bibfnamefont {Y.}~\bibnamefont {Kobayashi}}, \ and\
  \bibinfo {author} {\bibfnamefont {M.}~\bibnamefont {Sato}},\ }\href {\doibase
  10.1143/JPSJ.75.084707} {\bibfield  {journal} {\bibinfo  {journal} {J. Phys.
  Soc. Jpn.}\ }\textbf {\bibinfo {volume} {75}},\ \bibinfo {pages} {084707}
  (\bibinfo {year} {2006})}\BibitemShut {NoStop}%
\bibitem [{\citenamefont {Derakhshan}\ \emph {et~al.}(2007)\citenamefont
  {Derakhshan}, \citenamefont {Cuthbert}, \citenamefont {Greedan},
  \citenamefont {Rahaman},\ and\ \citenamefont {Saha-Dasgupta}}]{derakhshan07}%
  \BibitemOpen
  \bibfield  {author} {\bibinfo {author} {\bibfnamefont {S.}~\bibnamefont
  {Derakhshan}}, \bibinfo {author} {\bibfnamefont {H.~L.}\ \bibnamefont
  {Cuthbert}}, \bibinfo {author} {\bibfnamefont {J.~E.}\ \bibnamefont
  {Greedan}}, \bibinfo {author} {\bibfnamefont {B.}~\bibnamefont {Rahaman}}, \
  and\ \bibinfo {author} {\bibfnamefont {T.}~\bibnamefont {Saha-Dasgupta}},\
  }\href {\doibase 10.1103/PhysRevB.76.104403} {\bibfield  {journal} {\bibinfo
  {journal} {Phys. Rev. B}\ }\textbf {\bibinfo {volume} {76}},\ \bibinfo
  {pages} {104403} (\bibinfo {year} {2007})}\BibitemShut {NoStop}%
\bibitem [{\citenamefont {Yamanaka}\ \emph {et~al.}(1993)\citenamefont
  {Yamanaka}, \citenamefont {Hatsugai},\ and\ \citenamefont
  {Kohmoto}}]{AHC_phase_diagram}%
  \BibitemOpen
  \bibfield  {author} {\bibinfo {author} {\bibfnamefont {M.}~\bibnamefont
  {Yamanaka}}, \bibinfo {author} {\bibfnamefont {Y.}~\bibnamefont {Hatsugai}},
  \ and\ \bibinfo {author} {\bibfnamefont {M.}~\bibnamefont {Kohmoto}},\ }\href
  {\doibase 10.1103/PhysRevB.48.9555} {\bibfield  {journal} {\bibinfo
  {journal} {Phys. Rev. B}\ }\textbf {\bibinfo {volume} {48}},\ \bibinfo
  {pages} {9555} (\bibinfo {year} {1993})}\BibitemShut {NoStop}%
\bibitem [{\citenamefont {Koo}\ and\ \citenamefont {Whangbo}(2008)}]{koo08}%
  \BibitemOpen
  \bibfield  {author} {\bibinfo {author} {\bibfnamefont {H.-J.}\ \bibnamefont
  {Koo}}\ and\ \bibinfo {author} {\bibfnamefont {M.-H.}\ \bibnamefont
  {Whangbo}},\ }\href {\doibase 10.1021/ic701153z} {\bibfield  {journal}
  {\bibinfo  {journal} {Inorg. Chem.}\ }\textbf {\bibinfo {volume} {47}},\
  \bibinfo {pages} {128} (\bibinfo {year} {2008})}\BibitemShut {NoStop}%
\bibitem [{\citenamefont {Miura}\ \emph {et~al.}(2008)\citenamefont {Miura},
  \citenamefont {Yasui}, \citenamefont {Moyoshi}, \citenamefont {Sato},\ and\
  \citenamefont {Kakurai}}]{miura08}%
  \BibitemOpen
  \bibfield  {author} {\bibinfo {author} {\bibfnamefont {Y.}~\bibnamefont
  {Miura}}, \bibinfo {author} {\bibfnamefont {Y.}~\bibnamefont {Yasui}},
  \bibinfo {author} {\bibfnamefont {T.}~\bibnamefont {Moyoshi}}, \bibinfo
  {author} {\bibfnamefont {M.}~\bibnamefont {Sato}}, \ and\ \bibinfo {author}
  {\bibfnamefont {K.}~\bibnamefont {Kakurai}},\ }\href {\doibase
  10.1143/JPSJ.77.104709} {\bibfield  {journal} {\bibinfo  {journal} {J. Phys.
  Soc. Jpn.}\ }\textbf {\bibinfo {volume} {77}},\ \bibinfo {pages} {104709}
  (\bibinfo {year} {2008})}\BibitemShut {NoStop}%
\bibitem [{\citenamefont {Xu}\ \emph {et~al.}(2005)\citenamefont {Xu},
  \citenamefont {Assoud}, \citenamefont {Soheilnia}, \citenamefont
  {Derakhshan}, \citenamefont {Cuthbert}, \citenamefont {Greedan},
  \citenamefont {Whangbo},\ and\ \citenamefont {Kleinke}}]{xu05}%
  \BibitemOpen
  \bibfield  {author} {\bibinfo {author} {\bibfnamefont {J.}~\bibnamefont
  {Xu}}, \bibinfo {author} {\bibfnamefont {A.}~\bibnamefont {Assoud}}, \bibinfo
  {author} {\bibfnamefont {N.}~\bibnamefont {Soheilnia}}, \bibinfo {author}
  {\bibfnamefont {S.}~\bibnamefont {Derakhshan}}, \bibinfo {author}
  {\bibfnamefont {H.~L.}\ \bibnamefont {Cuthbert}}, \bibinfo {author}
  {\bibfnamefont {J.~E.}\ \bibnamefont {Greedan}}, \bibinfo {author}
  {\bibfnamefont {M.~H.}\ \bibnamefont {Whangbo}}, \ and\ \bibinfo {author}
  {\bibfnamefont {H.}~\bibnamefont {Kleinke}},\ }\href {\doibase
  10.1021/ic0502832} {\bibfield  {journal} {\bibinfo  {journal} {Inorg. Chem.}\
  }\textbf {\bibinfo {volume} {44}},\ \bibinfo {pages} {5042} (\bibinfo {year}
  {2005})}\BibitemShut {NoStop}%
\bibitem [{\citenamefont {Smirnova}\ \emph {et~al.}(2005)\citenamefont
  {Smirnova}, \citenamefont {Nalbandyan}, \citenamefont {Petrenko},\ and\
  \citenamefont {Avdeev}}]{smirnova05}%
  \BibitemOpen
  \bibfield  {author} {\bibinfo {author} {\bibfnamefont {O.~A.}\ \bibnamefont
  {Smirnova}}, \bibinfo {author} {\bibfnamefont {V.~B.}\ \bibnamefont
  {Nalbandyan}}, \bibinfo {author} {\bibfnamefont {A.~A.}\ \bibnamefont
  {Petrenko}}, \ and\ \bibinfo {author} {\bibfnamefont {M.}~\bibnamefont
  {Avdeev}},\ }\href {\doibase 10.1016/j.jssc.2005.02.002} {\bibfield
  {journal} {\bibinfo  {journal} {J. Solid State Chem.}\ }\textbf {\bibinfo
  {volume} {178}},\ \bibinfo {pages} {1165} (\bibinfo {year}
  {2005})}\BibitemShut {NoStop}%
\bibitem [{sup()}]{suppl}%
  \BibitemOpen
  \href@noop {} {}\bibinfo {note} {See Supplementary information for the
  crystal structures, full set of transfer integrals, depdence of the exchange
  integrals on the structural model, temperature dependence of magnetic
  susceptibility as a function of oxygen deficiency, as well as auxilary DOS
  and band structure plots.}\BibitemShut {Stop}%
\bibitem [{\citenamefont {Koepernik}\ and\ \citenamefont
  {Eschrig}(1999)}]{Koep1}%
  \BibitemOpen
  \bibfield  {author} {\bibinfo {author} {\bibfnamefont {K.}~\bibnamefont
  {Koepernik}}\ and\ \bibinfo {author} {\bibfnamefont {H.}~\bibnamefont
  {Eschrig}},\ }\href@noop {} {\bibfield  {journal} {\bibinfo  {journal} {Phys.
  Rev. B}\ }\textbf {\bibinfo {volume} {59}},\ \bibinfo {pages} {1743}
  (\bibinfo {year} {1999})}\BibitemShut {NoStop}%
\bibitem [{\citenamefont {Perdew}\ and\ \citenamefont {Wang}(1992)}]{PerdW}%
  \BibitemOpen
  \bibfield  {author} {\bibinfo {author} {\bibfnamefont {J.~P.}\ \bibnamefont
  {Perdew}}\ and\ \bibinfo {author} {\bibfnamefont {Y.}~\bibnamefont {Wang}},\
  }\href@noop {} {\bibfield  {journal} {\bibinfo  {journal} {Phys. Rev. B}\
  }\textbf {\bibinfo {volume} {45}},\ \bibinfo {pages} {13244} (\bibinfo {year}
  {1992})}\BibitemShut {NoStop}%
\bibitem [{\citenamefont {Kasinathan}\ \emph {et~al.}(2012)\citenamefont
  {Kasinathan}, \citenamefont {Wagner}, \citenamefont {Koepernik},
  \citenamefont {Cardoso-Gil}, \citenamefont {Grin},\ and\ \citenamefont
  {Rosner}}]{kasinathan2012}%
  \BibitemOpen
  \bibfield  {author} {\bibinfo {author} {\bibfnamefont {D.}~\bibnamefont
  {Kasinathan}}, \bibinfo {author} {\bibfnamefont {M.}~\bibnamefont {Wagner}},
  \bibinfo {author} {\bibfnamefont {K.}~\bibnamefont {Koepernik}}, \bibinfo
  {author} {\bibfnamefont {R.}~\bibnamefont {Cardoso-Gil}}, \bibinfo {author}
  {\bibfnamefont {Y.}~\bibnamefont {Grin}}, \ and\ \bibinfo {author}
  {\bibfnamefont {H.}~\bibnamefont {Rosner}},\ }\href {\doibase
  10.1103/PhysRevB.85.035207} {\bibfield  {journal} {\bibinfo  {journal} {Phys.
  Rev. B}\ }\textbf {\bibinfo {volume} {85}},\ \bibinfo {pages} {035207}
  (\bibinfo {year} {2012})}\BibitemShut {NoStop}%
\bibitem [{\citenamefont {Heyd}\ \emph {et~al.}(2003)\citenamefont {Heyd},
  \citenamefont {Scuseria},\ and\ \citenamefont {Ernzerhof}}]{HSE03}%
  \BibitemOpen
  \bibfield  {author} {\bibinfo {author} {\bibfnamefont {J.}~\bibnamefont
  {Heyd}}, \bibinfo {author} {\bibfnamefont {G.~E.}\ \bibnamefont {Scuseria}},
  \ and\ \bibinfo {author} {\bibfnamefont {M.}~\bibnamefont {Ernzerhof}},\
  }\href {\doibase 10.1063/1.1564060} {\bibfield  {journal} {\bibinfo
  {journal} {J. Chem. Phys.}\ }\textbf {\bibinfo {volume} {118}},\ \bibinfo
  {pages} {8207} (\bibinfo {year} {2003})}\BibitemShut {NoStop}%
\bibitem [{\citenamefont {Heyd}\ \emph {et~al.}(2006)\citenamefont {Heyd},
  \citenamefont {Scuseria},\ and\ \citenamefont {Ernzerhof}}]{HSE06}%
  \BibitemOpen
  \bibfield  {author} {\bibinfo {author} {\bibfnamefont {J.}~\bibnamefont
  {Heyd}}, \bibinfo {author} {\bibfnamefont {G.~E.}\ \bibnamefont {Scuseria}},
  \ and\ \bibinfo {author} {\bibfnamefont {M.}~\bibnamefont {Ernzerhof}},\
  }\href {\doibase 10.1063/1.2204597} {\bibfield  {journal} {\bibinfo
  {journal} {J. Chem. Phys.}\ }\textbf {\bibinfo {volume} {124}},\ \bibinfo
  {pages} {219906} (\bibinfo {year} {2006})}\BibitemShut {NoStop}%
\bibitem [{\citenamefont {Kresse}\ and\ \citenamefont
  {Furthm{\"{u}}ller}(1996{\natexlab{a}})}]{vasp1}%
  \BibitemOpen
  \bibfield  {author} {\bibinfo {author} {\bibfnamefont {G.}~\bibnamefont
  {Kresse}}\ and\ \bibinfo {author} {\bibfnamefont {J.}~\bibnamefont
  {Furthm{\"{u}}ller}},\ }\href {\doibase 10.1103/PhysRevB.54.11169} {\bibfield
   {journal} {\bibinfo  {journal} {Phys. Rev. B}\ }\textbf {\bibinfo {volume}
  {54}},\ \bibinfo {pages} {11169} (\bibinfo {year}
  {1996}{\natexlab{a}})}\BibitemShut {NoStop}%
\bibitem [{\citenamefont {Kresse}\ and\ \citenamefont
  {Furthm{\"{u}}ller}(1996{\natexlab{b}})}]{vasp2}%
  \BibitemOpen
  \bibfield  {author} {\bibinfo {author} {\bibfnamefont {G.}~\bibnamefont
  {Kresse}}\ and\ \bibinfo {author} {\bibfnamefont {J.}~\bibnamefont
  {Furthm{\"{u}}ller}},\ }\href {\doibase 10.1016/0927-0256(96)00008-0}
  {\bibfield  {journal} {\bibinfo  {journal} {Comput. Mater. Sci.}\ }\textbf
  {\bibinfo {volume} {6}},\ \bibinfo {pages} {15} (\bibinfo {year}
  {1996}{\natexlab{b}})}\BibitemShut {NoStop}%
\bibitem [{\citenamefont {Johnston}\ \emph {et~al.}(2000)\citenamefont
  {Johnston}, \citenamefont {Kremer}, \citenamefont {Troyer}, \citenamefont
  {Wang}, \citenamefont {Kl\"umper}, \citenamefont {Bud'ko}, \citenamefont
  {Panchula},\ and\ \citenamefont {Canfield}}]{HC_AHC_Johnston}%
  \BibitemOpen
  \bibfield  {author} {\bibinfo {author} {\bibfnamefont {D.~C.}\ \bibnamefont
  {Johnston}}, \bibinfo {author} {\bibfnamefont {R.~K.}\ \bibnamefont
  {Kremer}}, \bibinfo {author} {\bibfnamefont {M.}~\bibnamefont {Troyer}},
  \bibinfo {author} {\bibfnamefont {X.}~\bibnamefont {Wang}}, \bibinfo {author}
  {\bibfnamefont {A.}~\bibnamefont {Kl\"umper}}, \bibinfo {author}
  {\bibfnamefont {S.~L.}\ \bibnamefont {Bud'ko}}, \bibinfo {author}
  {\bibfnamefont {A.~F.}\ \bibnamefont {Panchula}}, \ and\ \bibinfo {author}
  {\bibfnamefont {P.~C.}\ \bibnamefont {Canfield}},\ }\href {\doibase
  10.1103/PhysRevB.61.9558} {\bibfield  {journal} {\bibinfo  {journal} {Phys.
  Rev. B}\ }\textbf {\bibinfo {volume} {61}},\ \bibinfo {pages} {9558}
  (\bibinfo {year} {2000})},\ \Eprint
  {http://arxiv.org/abs/arXiv:cond-mat/0003271} {arXiv:cond-mat/0003271}
  \BibitemShut {NoStop}%
\bibitem [{\citenamefont {Borras-Almenar}\ \emph {et~al.}(1994)\citenamefont
  {Borras-Almenar}, \citenamefont {Coronado}, \citenamefont {Curely},
  \citenamefont {Georges},\ and\ \citenamefont {Gianduzzo}}]{AHC_HTSE}%
  \BibitemOpen
  \bibfield  {author} {\bibinfo {author} {\bibfnamefont {J.~J.}\ \bibnamefont
  {Borras-Almenar}}, \bibinfo {author} {\bibfnamefont {E.}~\bibnamefont
  {Coronado}}, \bibinfo {author} {\bibfnamefont {J.}~\bibnamefont {Curely}},
  \bibinfo {author} {\bibfnamefont {R.}~\bibnamefont {Georges}}, \ and\
  \bibinfo {author} {\bibfnamefont {J.~C.}\ \bibnamefont {Gianduzzo}},\ }\href
  {\doibase 10.1021/ic00101a006} {\bibfield  {journal} {\bibinfo  {journal}
  {Inorg. Chem.}\ }\textbf {\bibinfo {volume} {33}},\ \bibinfo {pages} {5171}
  (\bibinfo {year} {1994})}\BibitemShut {NoStop}%
\bibitem [{\citenamefont {Todo}\ and\ \citenamefont {Kato}(2001)}]{looper}%
  \BibitemOpen
  \bibfield  {author} {\bibinfo {author} {\bibfnamefont {S.}~\bibnamefont
  {Todo}}\ and\ \bibinfo {author} {\bibfnamefont {K.}~\bibnamefont {Kato}},\
  }\href {\doibase 10.1103/PhysRevLett.87.047203} {\bibfield  {journal}
  {\bibinfo  {journal} {Phys. Rev. Lett.}\ }\textbf {\bibinfo {volume} {87}},\
  \bibinfo {pages} {047203} (\bibinfo {year} {2001})},\ \Eprint
  {http://arxiv.org/abs/cond-mat/9911047} {cond-mat/9911047} \BibitemShut
  {NoStop}%
\bibitem [{\citenamefont {Albuquerque}\ \emph {et~al.}(2007)\citenamefont
  {Albuquerque}, \citenamefont {Alet}, \citenamefont {Corboz}, \citenamefont
  {Dayal}, \citenamefont {Feiguin}, \citenamefont {Fuchs}, \citenamefont
  {Gamper}, \citenamefont {Gull}, \citenamefont {G\"urtler}, \citenamefont
  {Honecker}, \citenamefont {Igarashi}, \citenamefont {K\"orner}, \citenamefont
  {Kozhevnikov}, \citenamefont {L\"auchli}, \citenamefont {Manmana},
  \citenamefont {Matsumoto}, \citenamefont {McCulloch}, \citenamefont {Michel},
  \citenamefont {Noack}, \citenamefont {Pawlowski}, \citenamefont {Pollet},
  \citenamefont {Pruschke}, \citenamefont {Schollw\"ock}, \citenamefont {Todo},
  \citenamefont {Trebst}, \citenamefont {Troyer}, \citenamefont {Werner},\ and\
  \citenamefont {Wessel}}]{ALPS}%
  \BibitemOpen
  \bibfield  {author} {\bibinfo {author} {\bibfnamefont {A.}~\bibnamefont
  {Albuquerque}}, \bibinfo {author} {\bibfnamefont {F.}~\bibnamefont {Alet}},
  \bibinfo {author} {\bibfnamefont {P.}~\bibnamefont {Corboz}}, \bibinfo
  {author} {\bibfnamefont {P.}~\bibnamefont {Dayal}}, \bibinfo {author}
  {\bibfnamefont {A.}~\bibnamefont {Feiguin}}, \bibinfo {author} {\bibfnamefont
  {S.}~\bibnamefont {Fuchs}}, \bibinfo {author} {\bibfnamefont
  {L.}~\bibnamefont {Gamper}}, \bibinfo {author} {\bibfnamefont
  {E.}~\bibnamefont {Gull}}, \bibinfo {author} {\bibfnamefont {S.}~\bibnamefont
  {G\"urtler}}, \bibinfo {author} {\bibfnamefont {A.}~\bibnamefont {Honecker}},
  \bibinfo {author} {\bibfnamefont {R.}~\bibnamefont {Igarashi}}, \bibinfo
  {author} {\bibfnamefont {M.}~\bibnamefont {K\"orner}}, \bibinfo {author}
  {\bibfnamefont {A.}~\bibnamefont {Kozhevnikov}}, \bibinfo {author}
  {\bibfnamefont {A.}~\bibnamefont {L\"auchli}}, \bibinfo {author}
  {\bibfnamefont {S.~R.}\ \bibnamefont {Manmana}}, \bibinfo {author}
  {\bibfnamefont {M.}~\bibnamefont {Matsumoto}}, \bibinfo {author}
  {\bibfnamefont {I.~P.}\ \bibnamefont {McCulloch}}, \bibinfo {author}
  {\bibfnamefont {F.}~\bibnamefont {Michel}}, \bibinfo {author} {\bibfnamefont
  {R.~M.}\ \bibnamefont {Noack}}, \bibinfo {author} {\bibfnamefont
  {G.}~\bibnamefont {Pawlowski}}, \bibinfo {author} {\bibfnamefont
  {L.}~\bibnamefont {Pollet}}, \bibinfo {author} {\bibfnamefont
  {T.}~\bibnamefont {Pruschke}}, \bibinfo {author} {\bibfnamefont
  {U.}~\bibnamefont {Schollw\"ock}}, \bibinfo {author} {\bibfnamefont
  {S.}~\bibnamefont {Todo}}, \bibinfo {author} {\bibfnamefont {S.}~\bibnamefont
  {Trebst}}, \bibinfo {author} {\bibfnamefont {M.}~\bibnamefont {Troyer}},
  \bibinfo {author} {\bibfnamefont {P.}~\bibnamefont {Werner}}, \ and\ \bibinfo
  {author} {\bibfnamefont {S.}~\bibnamefont {Wessel}},\ }\href {\doibase
  10.1016/j.jmmm.2006.10.304} {\bibfield  {journal} {\bibinfo  {journal} {J.
  Magn. Magn. Mater.}\ }\textbf {\bibinfo {volume} {310}},\ \bibinfo {pages}
  {1187} (\bibinfo {year} {2007})},\ \Eprint
  {http://arxiv.org/abs/arXiv:0801.1765} {arXiv:0801.1765} \BibitemShut
  {NoStop}%
\bibitem [{\citenamefont {Schulenburg}()}]{spinpack}%
  \BibitemOpen
  \bibfield  {author} {\bibinfo {author} {\bibfnamefont {J.}~\bibnamefont
  {Schulenburg}},\ }\href@noop {} {}\bibinfo {note}
  {{\url{http://www-e.uni-magdeburg.de/jschulen/spin}}}\BibitemShut {NoStop}%
\bibitem [{\citenamefont {Belik}\ \emph {et~al.}(2004)\citenamefont {Belik},
  \citenamefont {Azuma},\ and\ \citenamefont
  {Takano}}]{HC_Sr2CuPO42_Ba2CuPO42_str_chiT_DTA}%
  \BibitemOpen
  \bibfield  {author} {\bibinfo {author} {\bibfnamefont {A.~A.}\ \bibnamefont
  {Belik}}, \bibinfo {author} {\bibfnamefont {M.}~\bibnamefont {Azuma}}, \ and\
  \bibinfo {author} {\bibfnamefont {M.}~\bibnamefont {Takano}},\ }\href
  {\doibase 10.1016/j.jssc.2003.09.024} {\bibfield  {journal} {\bibinfo
  {journal} {J. Solid State Chem.}\ }\textbf {\bibinfo {volume} {177}},\
  \bibinfo {pages} {883} (\bibinfo {year} {2004})}\BibitemShut {NoStop}%
\bibitem [{\citenamefont {Volkova}\ \emph {et~al.}(2010)\citenamefont
  {Volkova}, \citenamefont {Morozov}, \citenamefont {Shutov}, \citenamefont
  {Lapsheva}, \citenamefont {Sindzingre}, \citenamefont {C\'epas},
  \citenamefont {Yehia}, \citenamefont {Kataev}, \citenamefont {Klingeler},
  \citenamefont {B\"uchner},\ and\ \citenamefont
  {Vasiliev}}]{HC_NOCuNO33_chiT_CpT_ESR_simul}%
  \BibitemOpen
  \bibfield  {author} {\bibinfo {author} {\bibfnamefont {O.}~\bibnamefont
  {Volkova}}, \bibinfo {author} {\bibfnamefont {I.}~\bibnamefont {Morozov}},
  \bibinfo {author} {\bibfnamefont {V.}~\bibnamefont {Shutov}}, \bibinfo
  {author} {\bibfnamefont {E.}~\bibnamefont {Lapsheva}}, \bibinfo {author}
  {\bibfnamefont {P.}~\bibnamefont {Sindzingre}}, \bibinfo {author}
  {\bibfnamefont {O.}~\bibnamefont {C\'epas}}, \bibinfo {author} {\bibfnamefont
  {M.}~\bibnamefont {Yehia}}, \bibinfo {author} {\bibfnamefont
  {V.}~\bibnamefont {Kataev}}, \bibinfo {author} {\bibfnamefont
  {R.}~\bibnamefont {Klingeler}}, \bibinfo {author} {\bibfnamefont
  {B.}~\bibnamefont {B\"uchner}}, \ and\ \bibinfo {author} {\bibfnamefont
  {A.}~\bibnamefont {Vasiliev}},\ }\href {\doibase 10.1103/PhysRevB.82.054413}
  {\bibfield  {journal} {\bibinfo  {journal} {Phys. Rev. B}\ }\textbf {\bibinfo
  {volume} {82}},\ \bibinfo {pages} {054413} (\bibinfo {year} {2010})},\
  \Eprint {http://arxiv.org/abs/arXiv:1004.0444} {arXiv:1004.0444} \BibitemShut
  {NoStop}%
\bibitem [{\citenamefont {Hall}\ \emph {et~al.}(1981)\citenamefont {Hall},
  \citenamefont {Marsh}, \citenamefont {Weller},\ and\ \citenamefont
  {Hatfield}}]{AHC_HTSE_AFAF}%
  \BibitemOpen
  \bibfield  {author} {\bibinfo {author} {\bibfnamefont {J.~W.}\ \bibnamefont
  {Hall}}, \bibinfo {author} {\bibfnamefont {W.~E.}\ \bibnamefont {Marsh}},
  \bibinfo {author} {\bibfnamefont {R.~R.}\ \bibnamefont {Weller}}, \ and\
  \bibinfo {author} {\bibfnamefont {W.~E.}\ \bibnamefont {Hatfield}},\ }\href
  {\doibase 10.1021/ic50218a017} {\bibfield  {journal} {\bibinfo  {journal}
  {Inorg. Chem.}\ }\textbf {\bibinfo {volume} {20}},\ \bibinfo {pages} {1033}
  (\bibinfo {year} {1981})}\BibitemShut {NoStop}%
\bibitem [{\citenamefont {Schmitt}\ \emph {et~al.}(2009)\citenamefont
  {Schmitt}, \citenamefont {M{\'{a}}lek}, \citenamefont {Drechsler},\ and\
  \citenamefont {Rosner}}]{FHC_Li2ZrCuO4_DFT_simul}%
  \BibitemOpen
  \bibfield  {author} {\bibinfo {author} {\bibfnamefont {M.}~\bibnamefont
  {Schmitt}}, \bibinfo {author} {\bibfnamefont {J.}~\bibnamefont
  {M{\'{a}}lek}}, \bibinfo {author} {\bibfnamefont {S.-L.}\ \bibnamefont
  {Drechsler}}, \ and\ \bibinfo {author} {\bibfnamefont {H.}~\bibnamefont
  {Rosner}},\ }\href {\doibase 10.1103/PhysRevB.80.205111} {\bibfield
  {journal} {\bibinfo  {journal} {Phys. Rev. B}\ }\textbf {\bibinfo {volume}
  {80}},\ \bibinfo {pages} {205111} (\bibinfo {year} {2009})},\ \Eprint
  {http://arxiv.org/abs/arXiv:0911.0307} {arXiv:0911.0307} \BibitemShut
  {NoStop}%
\bibitem [{\citenamefont {Wolter}\ \emph {et~al.}(2012)\citenamefont {Wolter},
  \citenamefont {Lipps}, \citenamefont {Sch\"apers}, \citenamefont {Drechsler},
  \citenamefont {Nishimoto}, \citenamefont {Vogel}, \citenamefont {Kataev},
  \citenamefont {B\"uchner}, \citenamefont {Rosner}, \citenamefont {Schmitt},
  \citenamefont {Uhlarz}, \citenamefont {Skourski}, \citenamefont {Wosnitza},
  \citenamefont {S\"ullow},\ and\ \citenamefont {Rule}}]{linarite_2012}%
  \BibitemOpen
  \bibfield  {author} {\bibinfo {author} {\bibfnamefont {A.~U.~B.}\
  \bibnamefont {Wolter}}, \bibinfo {author} {\bibfnamefont {F.}~\bibnamefont
  {Lipps}}, \bibinfo {author} {\bibfnamefont {M.}~\bibnamefont {Sch\"apers}},
  \bibinfo {author} {\bibfnamefont {S.-L.}\ \bibnamefont {Drechsler}}, \bibinfo
  {author} {\bibfnamefont {S.}~\bibnamefont {Nishimoto}}, \bibinfo {author}
  {\bibfnamefont {R.}~\bibnamefont {Vogel}}, \bibinfo {author} {\bibfnamefont
  {V.}~\bibnamefont {Kataev}}, \bibinfo {author} {\bibfnamefont
  {B.}~\bibnamefont {B\"uchner}}, \bibinfo {author} {\bibfnamefont
  {H.}~\bibnamefont {Rosner}}, \bibinfo {author} {\bibfnamefont
  {M.}~\bibnamefont {Schmitt}}, \bibinfo {author} {\bibfnamefont
  {M.}~\bibnamefont {Uhlarz}}, \bibinfo {author} {\bibfnamefont
  {Y.}~\bibnamefont {Skourski}}, \bibinfo {author} {\bibfnamefont
  {J.}~\bibnamefont {Wosnitza}}, \bibinfo {author} {\bibfnamefont
  {S.}~\bibnamefont {S\"ullow}}, \ and\ \bibinfo {author} {\bibfnamefont
  {K.~C.}\ \bibnamefont {Rule}},\ }\href {\doibase 10.1103/PhysRevB.85.014407}
  {\bibfield  {journal} {\bibinfo  {journal} {Phys. Rev. B}\ }\textbf {\bibinfo
  {volume} {85}},\ \bibinfo {pages} {014407} (\bibinfo {year} {2012})},\
  \Eprint {http://arxiv.org/abs/arXiv:1110.4729} {arXiv:1110.4729} \BibitemShut
  {NoStop}%
\bibitem [{\citenamefont {Fuchs}\ \emph {et~al.}(2007)\citenamefont {Fuchs},
  \citenamefont {Furthm{\"u}ller}, \citenamefont {Bechstedt}, \citenamefont
  {Shishkin},\ and\ \citenamefont {Kresse}}]{2007_HSE03_PBE0_G0W0_Si_InN_ZnO}%
  \BibitemOpen
  \bibfield  {author} {\bibinfo {author} {\bibfnamefont {F.}~\bibnamefont
  {Fuchs}}, \bibinfo {author} {\bibfnamefont {J.}~\bibnamefont
  {Furthm{\"u}ller}}, \bibinfo {author} {\bibfnamefont {F.}~\bibnamefont
  {Bechstedt}}, \bibinfo {author} {\bibfnamefont {M.}~\bibnamefont {Shishkin}},
  \ and\ \bibinfo {author} {\bibfnamefont {G.}~\bibnamefont {Kresse}},\ }\href
  {\doibase 10.1103/PhysRevB.76.115109} {\bibfield  {journal} {\bibinfo
  {journal} {Phys. Rev. B}\ }\textbf {\bibinfo {volume} {76}},\ \bibinfo
  {pages} {115109} (\bibinfo {year} {2007})},\ \Eprint
  {http://arxiv.org/abs/cond-mat/0604447} {cond-mat/0604447} \BibitemShut
  {NoStop}%
\bibitem [{\citenamefont {Andersen}\ and\ \citenamefont
  {Saha-Dasgupta}(2000)}]{NMTO}%
  \BibitemOpen
  \bibfield  {author} {\bibinfo {author} {\bibfnamefont {O.~K.}\ \bibnamefont
  {Andersen}}\ and\ \bibinfo {author} {\bibfnamefont {T.}~\bibnamefont
  {Saha-Dasgupta}},\ }\href {\doibase 10.1103/PhysRevB.62.R16219} {\bibfield
  {journal} {\bibinfo  {journal} {Phys. Rev. B}\ }\textbf {\bibinfo {volume}
  {62}},\ \bibinfo {pages} {R16219} (\bibinfo {year} {2000})}\BibitemShut
  {NoStop}%
\bibitem [{\citenamefont {Andersen}\ and\ \citenamefont
  {Jepsen}(1984)}]{TB-LMTO-ASA}%
  \BibitemOpen
  \bibfield  {author} {\bibinfo {author} {\bibfnamefont {O.~K.}\ \bibnamefont
  {Andersen}}\ and\ \bibinfo {author} {\bibfnamefont {O.}~\bibnamefont
  {Jepsen}},\ }\href {\doibase 10.1103/PhysRevLett.53.2571} {\bibfield
  {journal} {\bibinfo  {journal} {Phys. Rev. Lett.}\ }\textbf {\bibinfo
  {volume} {53}},\ \bibinfo {pages} {2571} (\bibinfo {year}
  {1984})}\BibitemShut {NoStop}%
\bibitem [{\citenamefont {Rosner}\ \emph {et~al.}(2009)\citenamefont {Rosner},
  \citenamefont {Schmitt}, \citenamefont {Kasinathan}, \citenamefont {Ormeci},
  \citenamefont {Richter}, \citenamefont {Drechsler},\ and\ \citenamefont
  {Johannes}}]{SCPO_comment}%
  \BibitemOpen
  \bibfield  {author} {\bibinfo {author} {\bibfnamefont {H.}~\bibnamefont
  {Rosner}}, \bibinfo {author} {\bibfnamefont {M.}~\bibnamefont {Schmitt}},
  \bibinfo {author} {\bibfnamefont {D.}~\bibnamefont {Kasinathan}}, \bibinfo
  {author} {\bibfnamefont {A.}~\bibnamefont {Ormeci}}, \bibinfo {author}
  {\bibfnamefont {J.}~\bibnamefont {Richter}}, \bibinfo {author} {\bibfnamefont
  {S.-L.}\ \bibnamefont {Drechsler}}, \ and\ \bibinfo {author} {\bibfnamefont
  {M.~D.}\ \bibnamefont {Johannes}},\ }\href {\doibase
  10.1103/PhysRevB.79.127101} {\bibfield  {journal} {\bibinfo  {journal} {Phys.
  Rev. B}\ }\textbf {\bibinfo {volume} {79}},\ \bibinfo {pages} {127101}
  (\bibinfo {year} {2009})}\BibitemShut {NoStop}%
\bibitem [{\citenamefont {Ylvisaker}\ \emph {et~al.}(2009)\citenamefont
  {Ylvisaker}, \citenamefont {Pickett},\ and\ \citenamefont
  {Koepernik}}]{LSDA+U_Pickett}%
  \BibitemOpen
  \bibfield  {author} {\bibinfo {author} {\bibfnamefont {E.~R.}\ \bibnamefont
  {Ylvisaker}}, \bibinfo {author} {\bibfnamefont {W.~E.}\ \bibnamefont
  {Pickett}}, \ and\ \bibinfo {author} {\bibfnamefont {K.}~\bibnamefont
  {Koepernik}},\ }\href {\doibase 10.1103/PhysRevB.79.035103} {\bibfield
  {journal} {\bibinfo  {journal} {Phys. Rev. B}\ }\textbf {\bibinfo {volume}
  {79}},\ \bibinfo {pages} {035103} (\bibinfo {year} {2009})},\ \Eprint
  {http://arxiv.org/abs/arXiv:0808.1706} {arXiv:0808.1706} \BibitemShut
  {NoStop}%
\bibitem [{\citenamefont {Watanabe}\ and\ \citenamefont
  {Yokoyama}(1999)}]{AFHC_theory}%
  \BibitemOpen
  \bibfield  {author} {\bibinfo {author} {\bibfnamefont {S.}~\bibnamefont
  {Watanabe}}\ and\ \bibinfo {author} {\bibfnamefont {H.}~\bibnamefont
  {Yokoyama}},\ }\href {\doibase 10.1143/JPSJ.68.2073} {\bibfield  {journal}
  {\bibinfo  {journal} {J. Phys. Soc. Jpn.}\ }\textbf {\bibinfo {volume}
  {68}},\ \bibinfo {pages} {2073} (\bibinfo {year} {1999})},\ \Eprint
  {http://arxiv.org/abs/cond-mat/9902311} {cond-mat/9902311} \BibitemShut
  {NoStop}%
\bibitem [{\citenamefont {Hida}(1994)}]{AHC_Hida_FMAF_spectrum}%
  \BibitemOpen
  \bibfield  {author} {\bibinfo {author} {\bibfnamefont {K.}~\bibnamefont
  {Hida}},\ }\href {\doibase 10.1143/JPSJ.63.2514} {\bibfield  {journal}
  {\bibinfo  {journal} {J. Phys. Soc. Jpn.}\ }\textbf {\bibinfo {volume}
  {63}},\ \bibinfo {pages} {2514} (\bibinfo {year} {1994})}\BibitemShut
  {NoStop}%
\end{thebibliography}

\begin{thebibliography}{2}%
\makeatletter
\providecommand \@ifxundefined [1]{%
 \@ifx{#1\undefined}
}%
\providecommand \@ifnum [1]{%
 \ifnum #1\expandafter \@firstoftwo
 \else \expandafter \@secondoftwo
 \fi
}%
\providecommand \@ifx [1]{%
 \ifx #1\expandafter \@firstoftwo
 \else \expandafter \@secondoftwo
 \fi
}%
\providecommand \natexlab [1]{#1}%
\providecommand \enquote  [1]{``#1''}%
\providecommand \bibnamefont  [1]{#1}%
\providecommand \bibfnamefont [1]{#1}%
\providecommand \citenamefont [1]{#1}%
\providecommand \href@noop [0]{\@secondoftwo}%
\providecommand \href [0]{\begingroup \@sanitize@url \@href}%
\providecommand \@href[1]{\@@startlink{#1}\@@href}%
\providecommand \@@href[1]{\endgroup#1\@@endlink}%
\providecommand \@sanitize@url [0]{\catcode `\\12\catcode `\$12\catcode
  `\&12\catcode `\#12\catcode `\^12\catcode `\_12\catcode `\%12\relax}%
\providecommand \@@startlink[1]{}%
\providecommand \@@endlink[0]{}%
\providecommand \url  [0]{\begingroup\@sanitize@url \@url }%
\providecommand \@url [1]{\endgroup\@href {#1}{\urlprefix }}%
\providecommand \urlprefix  [0]{URL }%
\providecommand \Eprint [0]{\href }%
\providecommand \doibase [0]{http://dx.doi.org/}%
\providecommand \selectlanguage [0]{\@gobble}%
\providecommand \bibinfo  [0]{\@secondoftwo}%
\providecommand \bibfield  [0]{\@secondoftwo}%
\providecommand \translation [1]{[#1]}%
\providecommand \BibitemOpen [0]{}%
\providecommand \bibitemStop [0]{}%
\providecommand \bibitemNoStop [0]{.\EOS\space}%
\providecommand \EOS [0]{\spacefactor3000\relax}%
\providecommand \BibitemShut  [1]{\csname bibitem#1\endcsname}%
\let\auto@bib@innerbib\@empty
\bibitem [{\citenamefont {Smirnova}\ \emph {et~al.}(2005)\citenamefont
  {Smirnova}, \citenamefont {Nalbandyan}, \citenamefont {Petrenko},\ and\
  \citenamefont {Avdeev}}]{S:smirnova05}%
  \BibitemOpen
  \bibfield  {author} {\bibinfo {author} {\bibfnamefont {O.~A.}\ \bibnamefont
  {Smirnova}}, \bibinfo {author} {\bibfnamefont {V.~B.}\ \bibnamefont
  {Nalbandyan}}, \bibinfo {author} {\bibfnamefont {A.~A.}\ \bibnamefont
  {Petrenko}}, \ and\ \bibinfo {author} {\bibfnamefont {M.}~\bibnamefont
  {Avdeev}},\ }\href {\doibase 10.1016/j.jssc.2005.02.002} {\bibfield
  {journal} {\bibinfo  {journal} {J. Solid State Chem.}\ }\textbf {\bibinfo
  {volume} {178}},\ \bibinfo {pages} {1165} (\bibinfo {year}
  {2005})}\BibitemShut {NoStop}%
\bibitem [{\citenamefont {Xu}\ \emph {et~al.}(2005)\citenamefont {Xu},
  \citenamefont {Assoud}, \citenamefont {Soheilnia}, \citenamefont
  {Derakhshan}, \citenamefont {Cuthbert}, \citenamefont {Greedan},
  \citenamefont {Whangbo},\ and\ \citenamefont {Kleinke}}]{S:xu05}%
  \BibitemOpen
  \bibfield  {author} {\bibinfo {author} {\bibfnamefont {J.}~\bibnamefont
  {Xu}}, \bibinfo {author} {\bibfnamefont {A.}~\bibnamefont {Assoud}}, \bibinfo
  {author} {\bibfnamefont {N.}~\bibnamefont {Soheilnia}}, \bibinfo {author}
  {\bibfnamefont {S.}~\bibnamefont {Derakhshan}}, \bibinfo {author}
  {\bibfnamefont {H.~L.}\ \bibnamefont {Cuthbert}}, \bibinfo {author}
  {\bibfnamefont {J.~E.}\ \bibnamefont {Greedan}}, \bibinfo {author}
  {\bibfnamefont {M.~H.}\ \bibnamefont {Whangbo}}, \ and\ \bibinfo {author}
  {\bibfnamefont {H.}~\bibnamefont {Kleinke}},\ }\href {\doibase
  10.1021/ic0502832} {\bibfield  {journal} {\bibinfo  {journal} {Inorg. Chem.}\
  }\textbf {\bibinfo {volume} {44}},\ \bibinfo {pages} {5042} (\bibinfo {year}
  {2005})}\BibitemShut {NoStop}%
\end{thebibliography}
%

\begin{widetext}
\newpage
\begin{center}
{\large
Supplementary information for 
\smallskip

\textbf{Microscopic magnetic modeling for the $S$\,=\,$\frac12$ alternating chain compounds Na$_3$Cu$_2$SbO$_6$ and Na$_2$Cu$_2$TeO$_6$}
\medskip

\normalsize M. Schmitt, O. Janson, S. Golbs, M. Schmidt,

W. Schnelle, J. Richter, and H. Rosner}
\end{center}
\medskip

\begin{table}[htb]
\caption{Crystal structures for Na$_3$Cu$_2$SbO$_6$ used in the DFT study. The experimental structure
is adopted from Ref.~\onlinecite{S:smirnova05}.  The fictitious planar structures deviate from the experimental ones only by
the coordinate of the O$(2)$ site. The relaxed structures were optimized
with respect to the LDA total energy.}
\begin{ruledtabular}
\begin{tabular}{ccccccccc}
\multicolumn{9}{c}{Na$_3$Cu$_2$SbO$_6$-- experimental structures}\\ \hline
& \multicolumn{3}{c}{exp. } & & \multicolumn{3}{c}{relaxed} &  \\
& $x/a$ & $y/b$ & $z/c$ & & $x/a$ & $y/b$ & $z/c$ &   \\
Cu    & 0      & 0.6667 & 0                 & & 0   & 0.6667 & 0       & \\
Sb    & 0      & 0      & 0                       & & 0   & 0      & 0       & \\
O(1)  & 0.2931 & 0.3340 & 0.7750  & & -0.1987 & -0.1667 & -0.2234  & \\
O(2)  & 0.2404 & 0.5    & 0.1774     & & -0.2619 & 0    & 0.1734  & \\
Na(1) & 0      & 0.5    & 0.5                & & 0      & -0.5    & -0.5     & \\
Na(2) & 0.5    & 0.3280 & 0.5           & & 0    & -0.1732 & -0.5     & \\  \hline
\multicolumn{9}{c}{Na$_3$Cu$_2$SbO$_6$-- fictitious structures}\\ \hline
& \multicolumn{3}{c}{planar exp.} & & \multicolumn{3}{c}{planar relaxed} &  \\
O(2)  & 0.2931 & 0.5    & 0.7750   & & -0.1987 & 0.5    & -0.2234  &\\ 
\end{tabular}
\end{ruledtabular}
\end{table}

\begin{table}[htb]
\caption{Crystal structures for Na$_2$Cu$_2$TeO$_6$ used in the DFT study. The experimental structure
is adopted from Ref.~ \onlinecite{S:xu05}.  The fictitious planar structures deviate from the experimental ones only by
the coordinate of the O$(2)$ site. The relaxed structures were optimized
with respect to the LDA total energy.}
\begin{ruledtabular}
\begin{tabular}{ccccccccc}
\multicolumn{9}{c}{Na$_2$Cu$_2$TeO$_6$-- experimental structures}\\ \hline
& \multicolumn{3}{c}{exp.} & & \multicolumn{3}{c}{relaxed} & \\
& $x/a$ & $y/b$ & $z/c$ & & $x/a$ & $y/b$ & $z/c$ &  \\
Cu    & 0      & 0.66475 & 0               &  & 0       & -0.3353  & 0 &       \\
Te    & 0      & 0       & 0                       &  & 0       & 0       & 0 &      \\
O(1)  & 0.1936 & 0.1632  & 0.2121  && 0.1906 & 0.1682  & 0.2156 & \\
O(2)  & 0.7574 & 0     & 0.1640        & &  -0.2519 & 0     & 0.1648 & \\
Na    & 0      & 0.1839  & 0.5               & &0       & 0.1849  & -0.5   &  \\ \hline
\multicolumn{9}{c}{Na$_2$Cu$_2$TeO$_6$-- fictitious structures}\\ \hline
& \multicolumn{3}{c}{planar exp.} & & \multicolumn{3}{c}{planar relaxed} &  \\
O(2) & 0.1936 & 0.5     & 0.2121        & &  0.1906  & 0.5     & 0.2156 & \\
\end{tabular}
\end{ruledtabular}
\end{table}

\begin{table}[htb]
\caption{\label{S-tab_ts}Transfer integrals $t_{ij}$ (in meV) evaluated by
fitting the LSDA band structures for the different structural models:
experimental (exp), LSDA-relaxed (relaxed) and fictitious planar
structures originating from the experimental ones by shifting O(2) 
(planar exp and planar relaxed). For the notation of $t_{ij}$, see Fig.~5 in
the manuscript. ("$-$" $<1$\,meV)}
\begin{ruledtabular}
\begin{tabular}{l cccccccccc}
\multicolumn{11}{c}{Na$_3$Cu$_2$SbO$_6$}\\ \hline
$t_i$/meV & $t_{1a}$ & $t_{1b}$ & $t_{2}$ & $t^{ic}_{1a}$& $t^{ic}_{1b}$ & $t_{1a}^{il}$ & $t_{1b}^{il}$ &  $t_{0}^{d}$ & $t_{1}^{d}$ & $t_{0}^{a}$\\
 exp & {\bf 60.6} & {\bf 127} & 18.2& $-$27.8 & 17.0 & 5.8 & $-$6.6 & 21.8 & $-$4.6 & 17.4\\
relaxed & {\bf 68.2} & {\bf 134} & 18.1 & $-$32.3 & 20.6 & 6.4 &$-$7.2   & 20.9 & $-$3.8  & 19.2 \\
planar exp & {\bf 45.3} & {\bf 119} & 22.4 &  $-$7.8 & 9.4 & 14.1 & $-$18.8 & 30.1 & $-$13.2& $-$ \\ 
planar relaxed&  {\bf 55.6} & {\bf 125}  &23.8 & $-$9.2& 10.7 & 17.3&$-$21.7 &29.2 &$-$19.9 & $-$ \\
\hline
\multicolumn{11}{c}{Na$_2$Cu$_2$TeO$_6$}\\ \hline 
$t_i$/meV & $t_{1a}$ & $t_{1b}$ & $t_{2}$ & $t^{ic}_{1a}$& $t^{ic}_{1b}$ & $t_{1a}^{il}$ & $t_{1b}^{il}$ &  $t_{0}^{d}$ & $t_{1}^{d}$ & $t_{0}^{a}$\\
exp & {\bf 15.6} & {\bf 162} & 16.4& $-$38.5 & 24.7 & 2.8 &$-$ 2.5 & 13.7 & $-$ &25.5\\
relaxed & {\bf 42.5} & {\bf 152} & 17.3 & $-$42.4 & 26.3 & 3.4 &$-$3.9 & 14.5 &$-$ &23.1\\
planar exp. &{\bf 27.3} & {\bf 152} & 29.3& $-$12.6 & 12.4 & 14.1 & $-$16.0 & 25.6& $-$11.8 & 1.3\\
planar relaxed &{\bf 45.2} & {\bf 148} & 30.0& $-$12.8  & 12.7  &  16.0& $-$18.6 & 26.0 &$-$10.7  & $-$\\
\end{tabular}
\end{ruledtabular}
\end{table}


\begin{table}[htb]
\caption{\label{xxx} Exchange integrals derived from the LSDA+$U$ total energy
calculation for different structural models ($U_{3d}$\,=\,$6.0$\,eV). For comparison
the AFM parts of exchange  integrals estimated from the effective one-orbital
approach (tight-binding model $\rightarrow$ Hubbard model $\rightarrow$
Heisenberg model, $U_{\text{eff}}$\,=\,4\,eV) are given in brackets. The "$-$" sign
means that the respective coupling was not evaluated.}
\begin{ruledtabular}
\begin{tabular}{l ccccccc}
\multicolumn{8}{c}{Na$_3$Cu$_2$SbO$_6$}\\ \hline
$J_i$/meV & $J_{1a}$ & ($J_{1a}^{AFM}$) & $J_{1b}$ & ($J_{1b}^{AFM}$) & $J_{2}$+$J_{1b}^{ic}$+$J_{1b}^{il}$&   $J^{ic}_{1a}$+$J^{ic}_{1b}$& $J^{d}_{0}$+$J^{d}_{1}$ \\

 exp &  -11.6 & &12.9 & & -0.01& 0.8& 0.4 \\
  & & (3.7)& & (16.1) &  &  & \\ 
relaxed  &  -10.7 & & 13.9 & & 0.3  & 0.7  & $-$  \\
 & & (4.7) & & (18.0)  & &  &\\
planar exp & -19.7&  & 10.8 & & 0.9 & 0.2 & 1.2 \\
 & &  (2.1) & & (14.2) & & &  \\ 
planar relaxed& -19.5 & & 12.2 & & 1.0 & 0.2 & $-$ \\
 & &  (3.1) & & (15.7) & & & \\
\hline
\multicolumn{8}{c}{Na$_2$Cu$_2$TeO$_6$}\\ \hline 
$J_i$/meV & $J_{1a}$ &($J_{1a}^{AFM}$)  & $J_{1b}$ &  ($J_{1b}^{AFM}$) & $J_{2}$+$J_{1b}^{ic}$+$J_{1b}^{il}$ & $J^{ic}_{1a}$+$J^{ic}_{1b}$& $J^{d}_{0}$+$J^{d}_{1}$ \\
exp & -10.3 & & 19.9 & & 0.1 & 1.0 & 0.1  \\
 & & (0.2) & & (26.2) & & &  \\
relaxed & -10.5 & & 16.9 & & 0.2 & 1.0 & $-$ \\
 & & (1.8) & & (23.1) & & & \\
planar exp. &  -26.8  & & 18.2 & & 1.2 & 0.5 & 0.8 \\
 & & (0.8) & & (23.1) & & & \\
planar relaxed & -26.2& & 17.2& & 1.5 & 0.6  &  $-$\\
 & &  (2.0) & & (21.9) & & &  \\ 
 \end{tabular}
\end{ruledtabular}
\end{table}

\begin{table}[h]
\caption{\label{tesb_dist} Comparison of representative interatomic distances
(in~\r{A}) and bond angles (in deg.) for the different structures.
$d_{\text{NN}}$ and $d_{\text{NNN}}$ correspond to the nearest-neighbor and
next-nearest-neighbor Cu--Cu distances. Cu--O--Cu is the bond angle within the
structural dimer. O--O is the shortest distance between O atoms in the
neighboring structural dimers. Cu--O are the distances within the CuO$_4$
plaquette. $\delta$O is the distance (in\,\r{A}) between the O(2) atom in the
experimental structure and the respective atom in the ideal planar chain
geometry.}
\begin{ruledtabular}
\begin{tabular}{lcccccc}
structure & $d_{\text{NN}}$ & $d_{\text{NNN}}$ & Cu--O--Cu & O--O & Cu--O & $\delta$O \\
\hline \multicolumn{7}{c}{Na$_3$Cu$_2$SbO$_6$} \\
exp & 2.96 & 5.91& 95.27& 2.94 & 2.021/2.000& 0.39\\
planar exp & 2.96& 5.91& 94.22 & 2.94 & 2.021/2.017& 0\\
relaxed & 2.96 & 5.91 & 96.03 & 2.96 & 1.998/1.988 & 0.43\\
planar relaxed & 2.96 & 5.91 & 95.39 & 2.96 & 1.998/1.998& 0\\
\hline \multicolumn{7}{c}{Na$_2$Cu$_2$TeO$_6$} \\
exp & 2.86 & 5.82 & 91.27 & 2.83& 1.978/1.999& 0.55\\
planar exp & 2.86 & 5.82 & 95.48 & 2.83 & 1.978/1.931 & 0\\
relaxed & 2.86 & 5.82 & 92.91& 2.92& 1.950/1.972& 0.53\\
planar relaxed &2.86 & 5.82& 95.24& 2.92& 1.950/1.935 & 0\\
\end{tabular}
\end{ruledtabular}
\end{table}


\begin{figure}[h]
\includegraphics[width=.75\textwidth,angle=0]{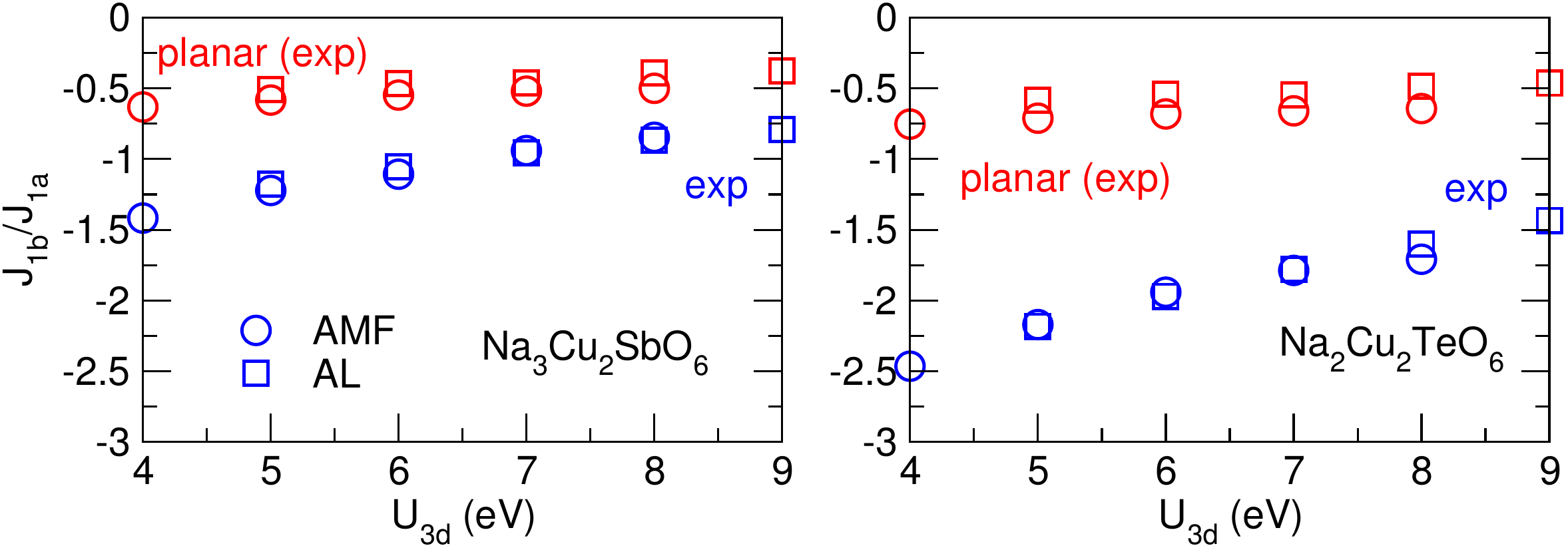}
\caption{\label{S-vgl_n2n3} 
Dependence of exchange integrals on the double-counting correction: the
around-mean-field (AMF) and the atomic limit (AL). For the LSDA+$U$
calculations we adopted the experimental crystal structure (exp) and the
structure with planar chains (planar exp).}
\end{figure}

\begin{figure}[h]
\includegraphics[width=.65\textwidth]{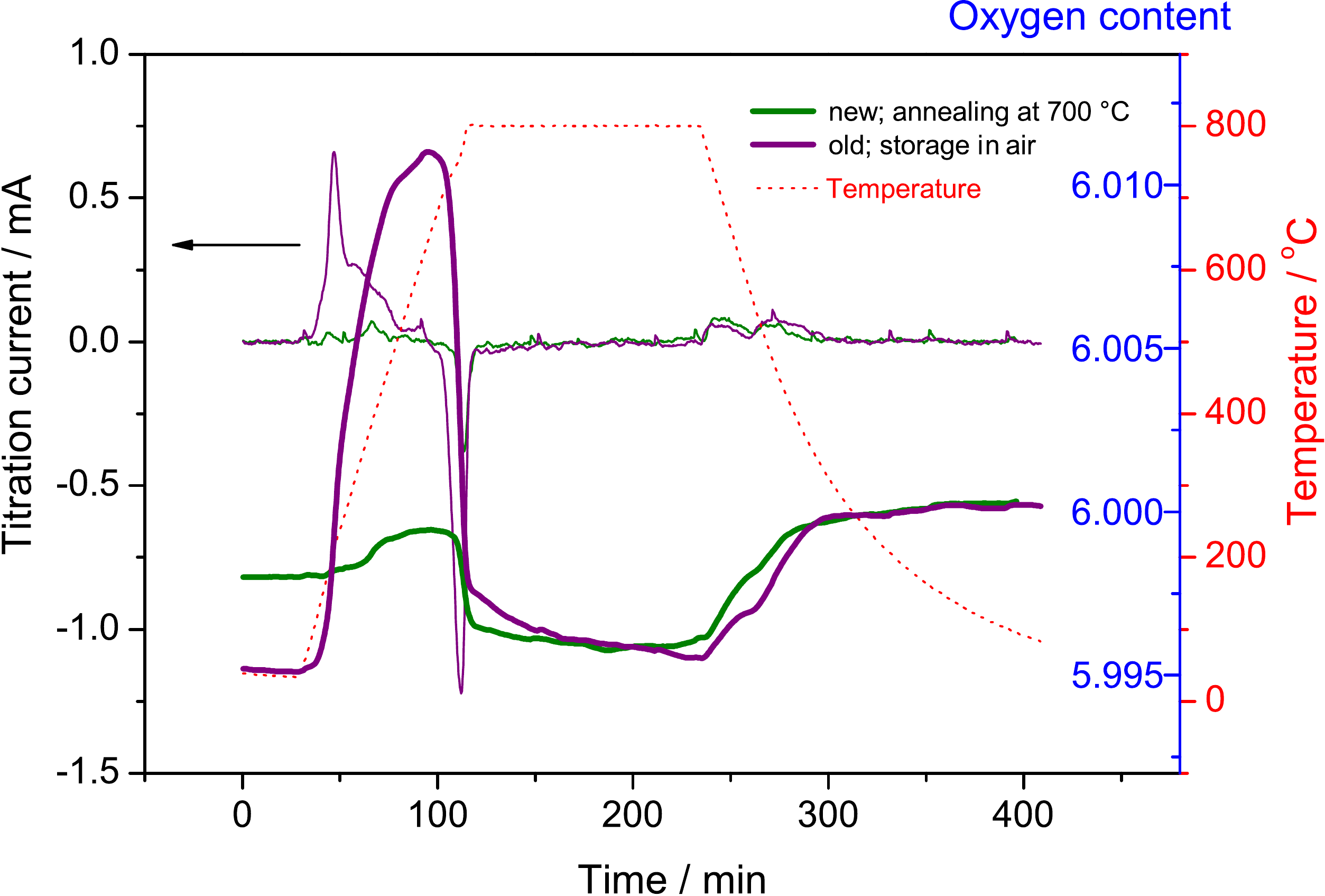}
\caption{\label{O_content} Oxygen content determined by coulometric titration
of a freshly annealed sample (green curve) and a sample stored at room
temperature and in air (violet curve).  Note the difference between the initial
oxygen content in the two samples. 
}
\end{figure}

\begin{figure}[h]
\includegraphics[width=.7\textwidth]{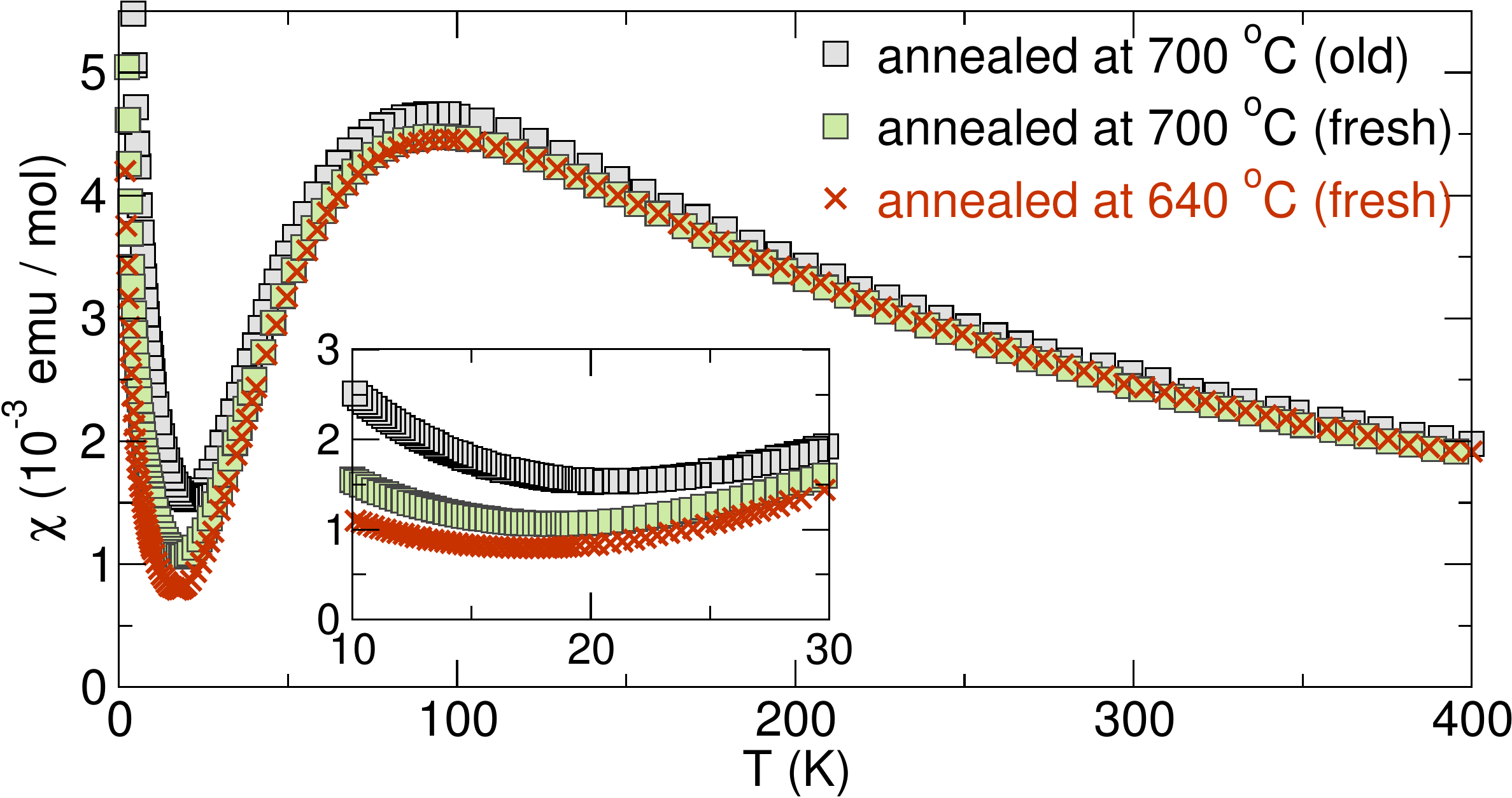}
\caption{\label{O_content_chi} 
Dependence of the magnetic susceptibility on the annealing temperature and
sample history (quality) due to changes in the O occupation.  External magnetic
field is 0.04\,T. Note the difference between the 700\,$^{\circ}$C samples
stored in air (``old'') and measured right after annealing (``fresh'').  In the
paper, we used $\chi(T)$ data for fresh samples annealed at 640\,$^{\circ}$C,
i.e., the samples featuring the smallest Curie-law-type impurities and/or defects (inset).}
\end{figure}

\begin{figure}[h]
\includegraphics[width=.5\textwidth,angle=0]{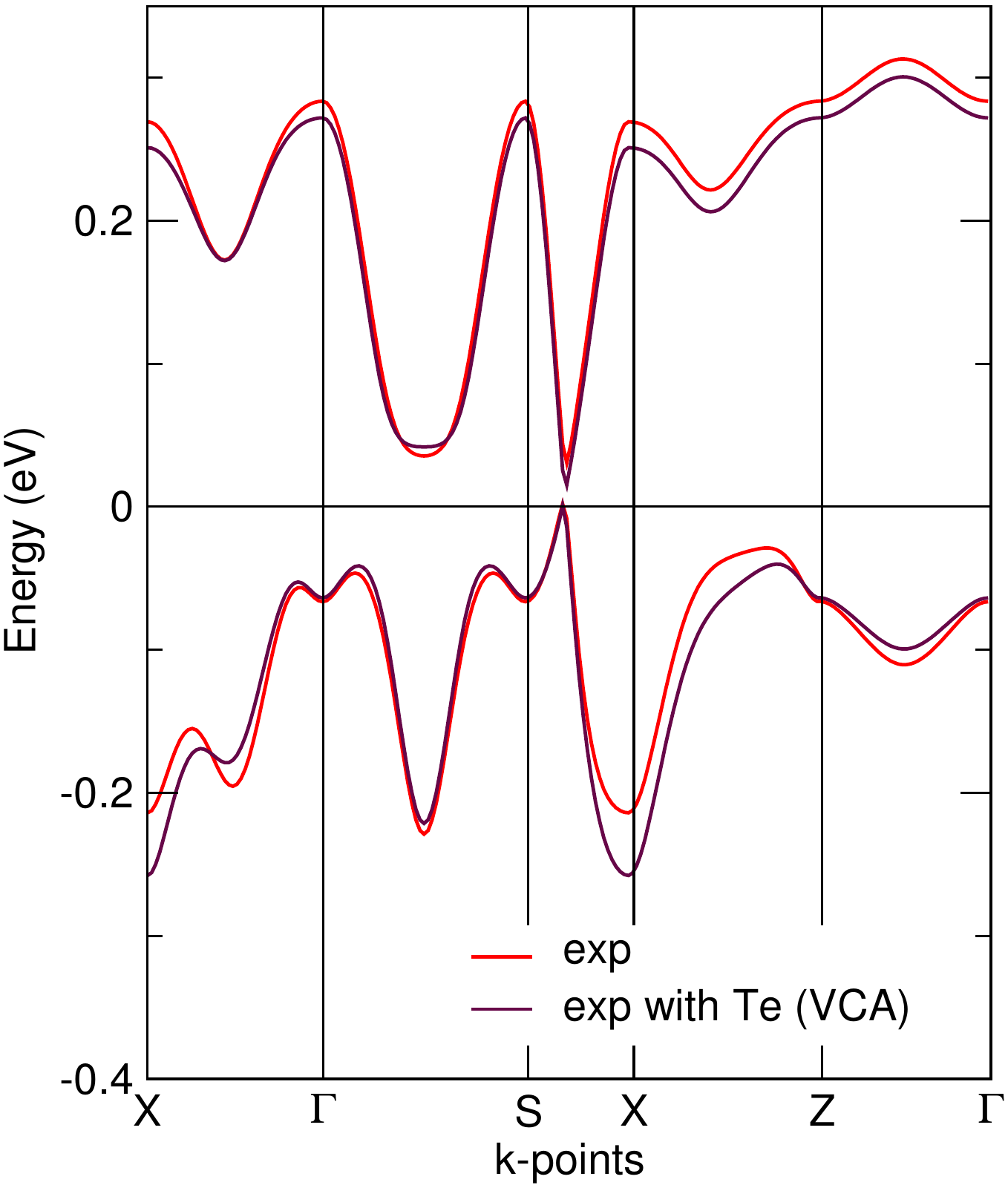}
\caption{\label{S-figS5} 
Comparison of the antibonding $dp\sigma$ band using LDA for the experimental
crystal structure of Na$_3$Cu$_2$SbO$_6$ and a model compound where Sb is
substituted by Te (with an concomitant change in the Na content) within the
same crystal structure using VCA. The obtained bands show nearly the same band
width and shape pointing to minor relevance of pure substitutional effects.}
\end{figure}

\end{widetext}

\end{document}